\documentclass[aps,prb,longbibliography,reprint,showpacs,superscriptaddress,amsmath]{revtex4-1}
\usepackage{color}
\usepackage{graphicx}
\usepackage{amsmath}

\begin{document}

%\title{Nematicity and C4-magnetism bordering orthorhombic magnetic order in FeSe under pressure}
%\title{Strengthened magnetic order within suppressed orthorhombicity in FeSe under pressure}
%\title{Distinct origins of coupled orthorhombicity and magnetic order in FeSe under pressure}
\title{Distinct pressure evolution of coupled nematic and magnetic order in FeSe}

%\title{Microscopic structural and magnetic investigation of FeSe under pressure}
%title{Pressure stabilizes magnetic order while suppressing orthorhombicity in FeSe}

%\author{Anna~E.~B\"ohmer }
%\affiliation{Ames Laboratory US DOE, Ames, Iowa 50011, USA}
%\author{Andreas Kreyssig }
%\affiliation{Ames Laboratory US DOE, Ames, Iowa 50011, USA}
\author{Anna E. B\"ohmer}
\thanks{These authors contributed equally.}
\affiliation{Ames Laboratory, U.S. DOE, Iowa State University, Ames, Iowa 50011, USA}
\affiliation{Institut f\"{u}r Festk\"{o}rperphysik, Karlsruhe Institute of Technology, 76021 Karlsruhe, Germany}

\author{Karunakar Kothapalli}
\thanks{These authors contributed equally.}
\affiliation{Ames Laboratory, U.S. DOE, Iowa State University, Ames, Iowa 50011, USA}
\affiliation{Department of Physics and Astronomy, Iowa State University, Ames, Iowa 50011, USA}
\affiliation{College of Arts and Sciences, King University, Bristol, Tennessee, 37620, USA}

\author{Wageesha T. Jayasekara}
\affiliation{Ames Laboratory, U.S. DOE, Iowa State University, Ames, Iowa 50011, USA}
\affiliation{Department of Physics and Astronomy, Iowa State University, Ames, Iowa 50011, USA}

\author{John M. Wilde}
\affiliation{Ames Laboratory, U.S. DOE, Iowa State University, Ames, Iowa 50011, USA}
\affiliation{Department of Physics and Astronomy, Iowa State University, Ames, Iowa 50011, USA}

\author{Bing Li}
\affiliation{Ames Laboratory, U.S. DOE, Iowa State University, Ames, Iowa 50011, USA}
\affiliation{Department of Physics and Astronomy, Iowa State University, Ames, Iowa 50011, USA}

\author{Aashish Sapkota}
\affiliation{Ames Laboratory, U.S. DOE, Iowa State University, Ames, Iowa 50011, USA}
\affiliation{Department of Physics and Astronomy, Iowa State University, Ames, Iowa 50011, USA}

\author{Benjamin G Ueland}
\affiliation{Ames Laboratory, U.S. DOE, Iowa State University, Ames, Iowa 50011, USA}
\affiliation{Department of Physics and Astronomy, Iowa State University, Ames, Iowa 50011, USA}

\author{Pinaki Das}
\affiliation{Ames Laboratory, U.S. DOE, Iowa State University, Ames, Iowa 50011, USA}
\affiliation{Department of Physics and Astronomy, Iowa State University, Ames, Iowa 50011, USA}

\author{Yumin Xiao}
\affiliation{HPCAT, Geophysical Laboratory, Carnegie Institution of Washington, Argonne, IL 60439, USA}

\author{Wenli Bi}
\affiliation{Advanced Photon Source, Argonne National Laboratory, Argonne, Illinois 60439, USA}
\affiliation{Department of Geology, University of Illinois at Urbana-Champaign, Urbana, Illinois 61801, USA}

\author{Jiyong Zhao}
\affiliation{Advanced Photon Source, Argonne National Laboratory, Argonne, Illinois 60439, USA}

\author{E. Ercan Alp}
\affiliation{Advanced Photon Source, Argonne National Laboratory, Argonne, Illinois 60439, USA}

\author{Sergey L. Bud'ko}
\affiliation{Ames Laboratory, U.S. DOE, Iowa State University, Ames, Iowa 50011, USA}
\affiliation{Department of Physics and Astronomy, Iowa State University, Ames, Iowa 50011, USA}

\author{Paul C. Canfield}
\affiliation{Ames Laboratory, U.S. DOE, Iowa State University, Ames, Iowa 50011, USA}
\affiliation{Department of Physics and Astronomy, Iowa State University, Ames, Iowa 50011, USA}

\author{Alan I. Goldman}
\affiliation{Ames Laboratory, U.S. DOE, Iowa State University, Ames, Iowa 50011, USA}
\affiliation{Department of Physics and Astronomy, Iowa State University, Ames, Iowa 50011, USA}

\author{Andreas Kreyssig}
\affiliation{Ames Laboratory, U.S. DOE, Iowa State University, Ames, Iowa 50011, USA}
\affiliation{Department of Physics and Astronomy, Iowa State University, Ames, Iowa 50011, USA}

\begin{abstract}
FeSe, despite being the structurally simplest compound in the family of iron-based superconductors, shows an astoundingly rich interplay of physical phenomena including nematicity and pressure-induced magnetism. Here, we present a microscopic study of these two phenomena by high-energy x-ray diffraction and time-domain M\"ossbauer spectroscopy on FeSe single crystals over a wide temperature and pressure range. The topology of the pressure-temperature phase diagram is a surprisingly close parallel to the well-known doping-temperature phase diagram of BaFe$_2$As$_2$ generated through partial Fe/Co and Ba/Na substitution. In FeSe with pressure $p$ as a control parameter, the magneto-structural ground state can be tuned from "pure" nematic --- paramagnetic with an orthorhombic lattice distortion --- through a strongly coupled magnetically ordered and orthorhombic state to a magnetically ordered state without an orthorhombic lattice distortion.  The magnetic hyperfine field increases monotonically over a wide pressure range. However, the orthorhombic distortion initially decreases under increasing pressure, but is stabilized by cooperative coupling to the pressure-induced magnetic order. Close to the reported maximum of the superconducting critical temperature $T_c$ (occuring at $p=6.8$ GPa), the orthorhombic distortion suddenly disappears and FeSe remains tetragonal down to the lowest temperature measured. Analysis of the structural and magnetic order parameters suggests an independent origin of the structural and magnetic ordering phenomena, and their cooperative coupling leads to the similarity with the canonical phase diagram of iron pnictides. 
\end{abstract}

\maketitle 
		
		\section{Introduction}
        A fascinating characteristic of iron-based superconductors is their complex phase diagrams, and a decade of research has revealed intricate relationships between their magnetism, structure and superconductivity. Most parent compounds of the iron-based superconductors support an antiferromagnetic ground state, and, similar to many other unconventional superconductors, the antiferromagnetic order needs to be sufficiently suppressed for superconductivity to occur. The antiferromagnetic order in the iron-based compounds is typically stripe-type, characterized by a wavevector that breaks the tetragonal lattice symmetry\cite{Dai15_review}. Hence, when stripe-type magnetic order forms, the lattice necessarily distorts from its high-temperature tetragonal structure and becomes orthorhombic\cite{Cano2010}. A dome of superconductivity arises around the point at which the magnetic order and lattice distortion are suppressed by a tuning parameter like doping or pressure, and the fluctuations related to the suppressed magnetic order and lattice distortion are a promising candidate for the superconducting pairing glue. Even seemingly special cases like the 10-3-8 material Ca$_{10}$(Pt$_3$As$_8$)(Fe$_2$As$_2$)$_5$ with a more complex chemical structure and disorder essentially fit into this picture\cite{Sapkota2014}.
        	
		There are two intriguing exceptions to the intimate relationship between the orthorhombic distortion and stripe-type magnetism, both of which have separately generated enormous interest. First, it was observed soon after the discovery of superconductivity in Fe-based materials that the orthorhombic lattice distortion may decouple from the stripe-type magnetic order and occur at a higher temperature ($T_s>T_N$)\cite{Cruz2008,Ni2008,Chu2009}. This has sparked the idea that the structural distortion is related to an independent "nematic" degree of freedom\cite{Fernandes2014} that could, in principle, exist without the magnetic order. Nevertheless, the orthorhombic distortion has theoretically been shown to be a likely consequence of stripe-type magnetic fluctuations\cite{Fernandes2010,Nandi2010,Fernandes2014}. Such split transitions are indeed very common and observed in pure and underdoped 1111-type materials\cite{Cruz2008}, in transition-metal substituted BaFe$_2$As$_2$\cite{Ni2010} and SrFe$_2$As$_2$, and in Co-substituted NaFeAs\cite{Parker2010}. 
        
		A few years later, it was discovered that magnetic order in certain iron-based systems can, in fact, also occur without an orthorhombic lattice distortion \cite{Hassinger2012,Avci2014,Boehmer2015II}.  These intriguing "C4-type", tetragonal magnetic phases were shown to arise from a coherent superposition of the two symmetry-equivalent, stripe-type antiferromagnetic wavevectors\cite{Allred2016}. The occurence of magnetic order within a tetragonal structure is almost ubiquitous to hole-doped 122-type systems, occurring in Ba(Fe,Mn)$_2$As$_2$\cite{Kreyssig2010}, (Ba,K)Fe$_2$As$_2$\cite{Hassinger2012,Boehmer2015II,Taddei2015,Hassinger_2016}, (Ba,Na)Fe$_2$As$_2$\cite{Avci2014,Wang2017II}, (Sr,Na)Fe$_2$As$_2$\cite{Allred2016,Taddei2016} and  (Ca,Na)Fe$_2$As$_2$\cite{Taddei2017}, and was recently also shown in effectively hole-doped CaK(Fe,Ni)$_4$As$_4$\cite{Meier2018}. 
		
		FeSe has generated enormous interest over the past few years as an extreme case of nematicity\cite{McQueen2009,Watson2015,Wang2015,Fanfarillo2016,Yamakawa_PRX_16,Tanatar2016,Chinotti2017,Watson2017II,He2017,Boehmer2017}. At ambient pressure, FeSe exhibits a tetragonal-to-orthorhombic transition close to $T_s=90$ K and no magnetic order down to sub-Kelvin temperatures\cite{Bendele2010}. In this sense, FeSe is the iron-based material with the largest extent of a purely nematic phase. FeSe exhibits a complex magnetic fluctuation spectrum with intensity occurring around both the stripe-type and N\'eel-type wavevectors\cite{Wang2015}, and the spectral weights shift in favor of the stripe-type fluctuations below $T_s=90$ K \cite{Wang2015GS}. A second reason why FeSe has generated excitement is the high tunability of its superconducting transition temperature $T_c$. Whereas ambient-pressure FeSe is superconducting below a modest $T_c$ of $\sim8$ K\cite{Hsu2008}, $T_c$ quadruples to $37$ K under hydrostatic pressure\cite{Mizuguchi2008,Medvedev2009,Margadonna2009,Garbarino2009}. Studies of monolayer films of FeSe grown on SrTiO$_3$ have even shown evidence for $T_c>100$ K (Ref. \onlinecite{Ge2015}). 
        
 \begin{figure}
			\includegraphics[width=8.6cm]{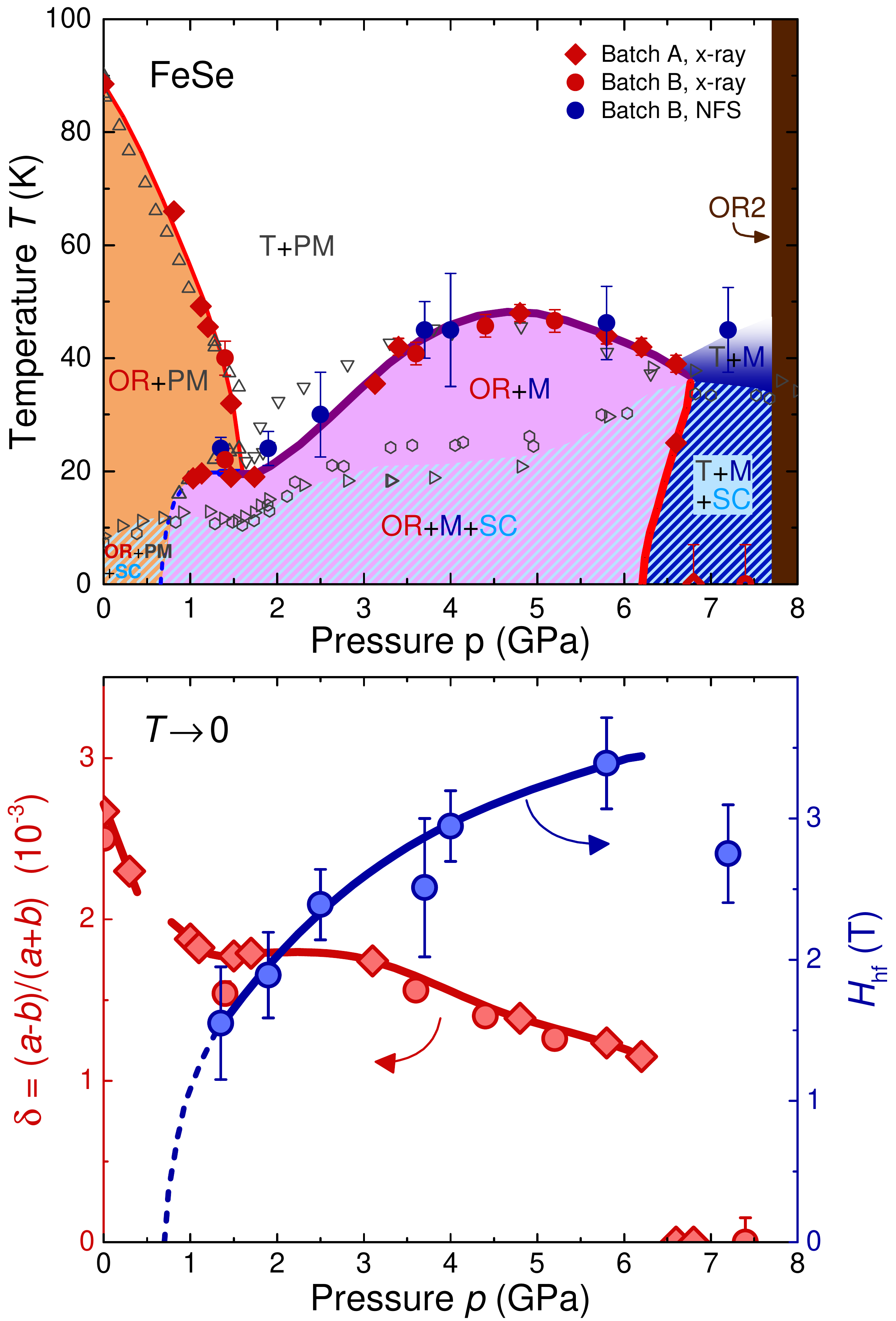}
			\caption{(a) Temperature-pressure phase diagram of bulk FeSe. The structural transition temperatures were obtained by x-ray diffraction measurements on two different batches: A (red circles) and B (red diamonds). T (OR) stands for the tetragonal (orthorhombic) phase and OR2 stands for the reduced-volume high-pressure orthorhombic phase. SC denotes the superconducting state. PM (M) indicates paramagnetic (magnetically ordered) regions of the phase diagram as determined from nuclear-forward-scattering experiments on samples from batch B, with the transition temperature indicated by blue circles. Thick (thin) lines represent first (second) order phase transitions, respectively. Data from other reports are shown by open gray symbols for comparison and completeness: up triangles\cite{Kaluarachchi2016} and down triangles\cite{Sun_2016} indicate the structural and magnetostructural transitions, right triangles\cite{Sun_2016} and hexagons\cite{Miyoshi2014} indicate the superconducting transition. (b) Pressure-dependence of the zero-temperature limit of the orthorhombic distortion, i.e. the structural order parameter (left axis), and the ordered hyperfine field (right axis), as a proxy for the magnetic order parameter.}	
			\label{fig:6}
		\end{figure}
        
		The temperature-pressure phase diagram of FeSe has been uncovered only incrementally. The structural transition is suppressed under pressure\cite{Miyoshi2014,Terashima2015,Kothapalli2016,Wang2016}, pressure-induced magnetic order was demonstrated for $p>0.8$  GPa and $T_N$ increases with further increasing pressure \cite{Bendele2010,Bendele2012,Terashima2015,Sun_2016,Kothapalli2016}. This, at first glance, represents a clear difference between FeSe and "typical" iron arsenides\cite{Buechner2009}.
		Under pressure, the structural and magnetic phase lines for FeSe in fact merge into a concomitant magneto-structural transition\cite{Kothapalli2016}, indicating that the magnetic ground state is orthorhombic and presumably the same stripe-type as occurs in other iron-based materials. The occurrence of stripe-type magnetic order is also suggested by NMR and muSR results\cite{Wang2016,Khasanov2016}. The magneto-structural transition temperature traces out a dome with increasing pressure, with a maximum of $45$ K occurring around 5 GPa \cite{Sun_2016}.
		
		 Here, we present a microscopic study of the magnetism, crystallographic symmetry, and in-plane lattice parameters of FeSe under hydrostatic pressure. The temperature dependence of the ordered magnetic hyperfine field and of the orthorhombic distortion in vapor-grown single crystals have been determined over a pressure range of $p=0$--$10$ GPa using high-energy x-ray diffraction and nuclear forward scattering. Figure 1 summarizes the experimental results.         
        In the temperature-pressure phase diagram of FeSe in Fig. 1(a), the nematic, tetragonal-paramagnetic (T+PM) region on the low-pressure side borders an orthorhombic-magnetically-ordered dome (OR+M). We find that magnetic order persists on the high-pressure side of the magnetic dome in the absence of an orthorhombic distortion (T+M). The whole phase diagram is "cut-off" by a sharp and first-order structural transition at $p=7.7$ GPa into an orthorhombic "OR2" phase \cite{Kumar2010,Svitlyk2016}. The low-temperature values of the magnetic and structural order parameters are shown in Fig. 1(b). The ordered magnetic hyperfine field increases monotonically over the orthorhombic-magnetic dome, whereas the orthorhombic distortion has a complex pressure dependence and mainly decreases on increasing pressure. The temperature-pressure phase diagram of FeSe is a surprisingly close parallel to the temperature-doping phase diagram of the $122$-type systems, if both electron-and hole-doping are considered. On the other hand, an analysis of the pressure dependent orthorhombic distortion and magnetic hyperfine field suggests that these ordering phenomena have distinct origins, although the order parameters couple cooperatively. 
			
		\section{Experimental Methods}
Single-crystals of FeSe were prepared by chemical vapor transport as described in Ref. \onlinecite{Boehmer2016II}. Batch A samples are from several batches using natural-abundance elements, whereas batch B samples are from a batch prepared using $95\%$ enriched $^{57}$Fe. As described in Ref. \onlinecite{Boehmer2016II}, the sample properties can vary even with tiny variations in growth conditions. Batch B was found to have less perfect mosaicity and less sharp phase transitions than samples from batch A, however batch A and B both have very similar transition temperatures and values for the orthorhombic distortion (see Fig.\ \ref{fig:6}).

High-energy ($100$ keV), high-resolution x-ray diffraction experiments were performed at endstation $6$ID-D of the APS Argonne on samples from batches A and B. The samples were pressurized in diamond anvil cells (DACs) using He gas as a highly hydrostatic pressure-transmitting medium. We used diamonds with 600 $\mu$m culets and stainless-steel and CuBe gaskets preindented to thicknesses of $\sim60$ $\mu$m and with laser-drilled holes of $\sim250-350$ $\mu$m. The position of a fluorescence line for ruby was used for ambient-temperature pressure calibration. Measurements of the lattice parameter of polycrystalline silver were used for in-situ pressure determination at all temperatures, so that the actual temperature-dependent pressure values are reported. Extended regions of selected reciprocal lattice planes and the powder diffraction pattern of silver were recorded by a MAR$345$ image plate system positioned $1.474$ m behind the DAC, as the DAC was rocked by up to $\pm 3.2^\circ$ about two independent axes perpendicular to the incident x-ray beam. High-resolution diffraction patterns of selected Bragg reflections of samples from batch A were also recorded using a Pixirad-$1$ detector positioned $1.397$ m behind the DAC while rocking around one of the two axes perpendicular to the x-ray beam. The in-plane lattice parameters were determined by fitting the Bragg peak positions after integrating the data over the transverse scattering directions. This procedure was used for both the data recorded by the Pixirad-$1$ detector and by the MAR$345$ image plate system. 

Nuclear forward scattering (NFS), i.e. time-domain M\"ossbauer spectroscopy, was performed on stations $3$ID-B and $16$ID-D at the APS on samples from batch B. Diamond anvil cells with He as a pressure transmitting medium and ruby as an in-situ pressure calibrant were used and the pressure cells were set up in a similar way as for the diffraction experiments. At 3ID-B, miniature panoramic DACs \cite{Bi2015} were used. The incident x-ray beam was monochromated to the $^{57}$Fe nuclear resonance energy of $14.4125$ keV with a resolution of 2  meV and the intensity of the scattered beam in the forward direction was recorded by an Avalanche Photo Diode detector. The beam size at $3$ID-B and $16$ID-D was $10\times10$ $\mu$m$^2$ and $20\times30$ $\mu$m$^2$, respectively. 
Spectra for $p\ge 2.5$ GPa were collected at $16$ID-D using the 24-bunch standard timing mode of the Advanced Photon Source, where an x-ray pulse of 80 ps duration hits the sample with a periodicity of 153 ns. Spectra for $p< 2.5$ GPa were collected at 3ID-B beamline in the so-called “hybrid mode” with a $1.5$ $\mu$s clear time for measurements after the initial excitation pulse hits the nuclei. This long-pulse mode reduces the counting rate by an order of magnitude. However, it drastically improves the sensitivity and precision of determining the internal magnetic hyperfine field due to the increased observation time of the nuclear decay, which is particularly relevant when hyperfine fields are very small. 
%Spectra for  were collected at $16$ID-D with a pulse separation of $153$ ns. Data for  GPa were collected at $3$ID-B utilizing the $1.2~\mu$s pulse separation available at the station. The longer pulse separation allows for observation of the quantum beats at longer delay times, which increases the sensitivity to very small magnetic hyperfine fields. 
The program \textsc{conuss} \cite{Sturhahn2000} was used to analyze the spectra and determine the  magnetic field hyperfine at the iron sites. 

\section{Results}
\subsection{Nuclear forward scattering and x-ray diffraction}

\begin{figure}
	\includegraphics[width=8.6cm]{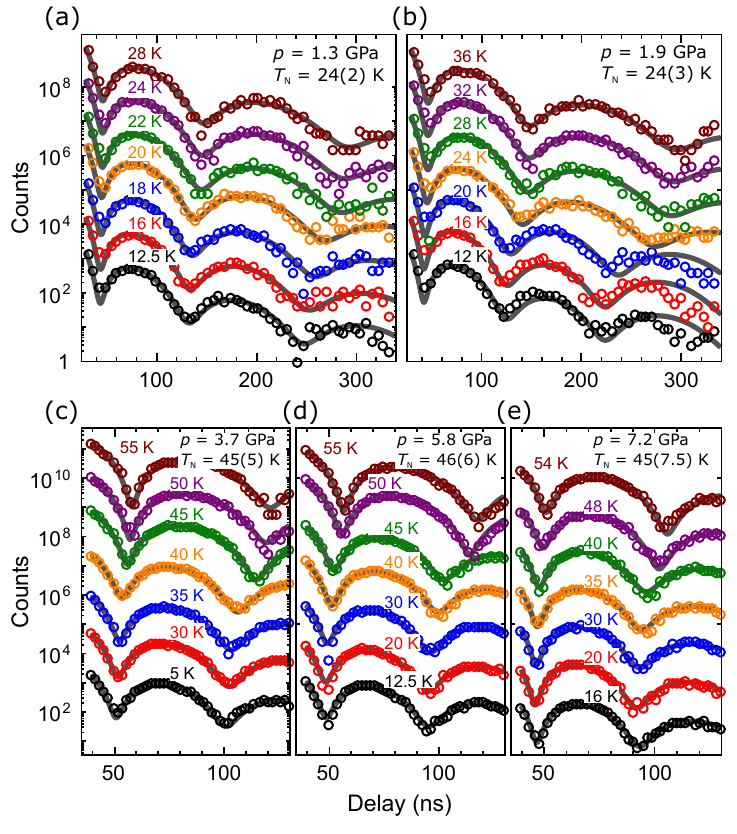}
	\caption{$^{57}$Fe nuclear-forward-scattering spectra for FeSe at selected pressures and temperatures. Data are offset for clarity, and dark gray lines show fits to the data using \textsc{conuss}. Data in (a) and (b) were collected on a $6~\mu$m thick sample from batch B in the long-pulse mode with a $1.5$ $\mu$s clear time for measurements. Data in (c)--(e) were collected on an $18~\mu$m thick sample from batch B using the 24 bunch standard timing mode of the APS, with a 153 ns separation between x-ray pulses.}
	\label{fig:2}
\end{figure}

\begin{figure*}
	\includegraphics[width=17.2cm]{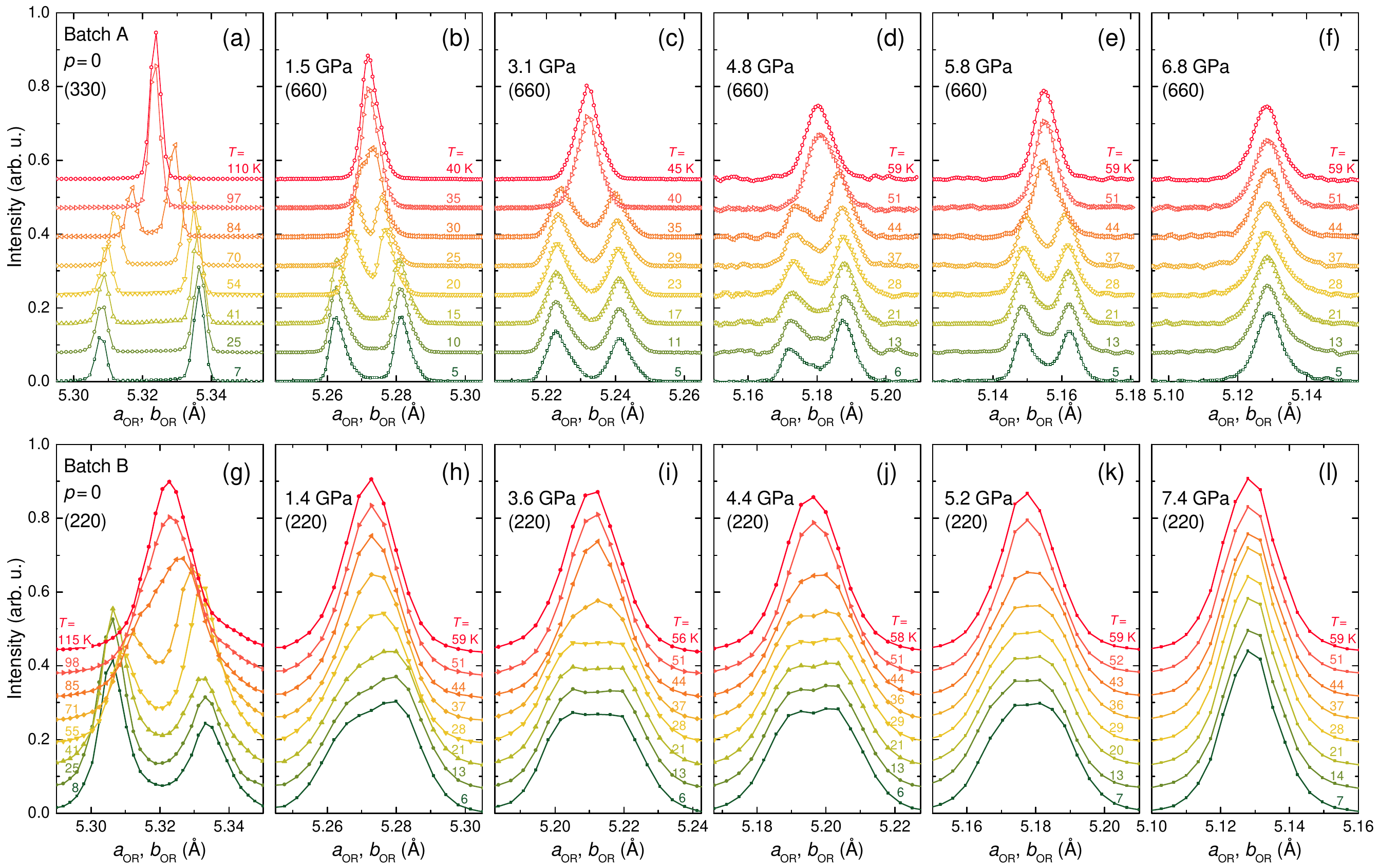}
	\caption{High-energy x-ray diffraction patterns demonstrating the tetragonal to orthorhombic phase transition.  $a_\mathrm{OR}$ and $b_\mathrm{OR}$ are the lattice parameters for the orthorhombic unit cell  (a)-(f), X-ray intensity profiles taken across the $(3~3~0)$ (a) and $(6~6~0)$ [(b)--(f)] tetragonal unit cell Bragg peaks on samples from batch A for various pressures and temperatures. The peak splitting results from the orthorhombic distortion. (g)--(l), X-ray intensity profiles close to the $(2~2~0)$ Bragg peaks for samples from batch B at various pressures and temperatures. Here, the peak splitting or broadening signals the orthorhombic distortion. The difference in peak profiles with respect to the upper panels results from the lower order of the chosen Bragg peak, the use of a different detector, and the broader mosaicity of samples from batch B.}	
	\label{fig:2b}
\end{figure*}

Figure \ref{fig:2} shows the NFS spectra at various pressures up to $7.2$ GPa, from which the information about the magnetic order is obtained. Data for low pressures were obtained to longer delay times on a $6~\mu$m thick sample so that small magnetic hyperfine fields could be determined more accurately. Data for pressures above $2.5$ GPa were obtained on an $18~\mu$m thick sample. The observed quantum beats originate from a convolution of the hyperfine field, quadrupole splitting and sample thickness. A change in the spectra, most notably the shift of the minima, e.g., between $20$ K and $24$ K at $1.9$ GPa, indicates that a magnetic phase transition has occurred. Such a transition is observed for pressures up to $7.2$ GPa. At 1.3 GPa a similar, though more continuous shift is discernible and identified as a magnetic phase transition. The corresponding magnetic hyperfine fields are reported in Fig.\ \ref{fig:3} below.  %A continuous magnetic phase transition is observed at 1.3 GPa, and more abruptly at 1.9 GPa. The subtle shifts are results of a very small hyperfine field. At the higher pressures, a more abrupt change of the spectra indicates a first-order magnetic phase transition. 

%\subsection{Orthorhombic distortion}

\begin{figure*}
	\includegraphics[width=17.2cm]{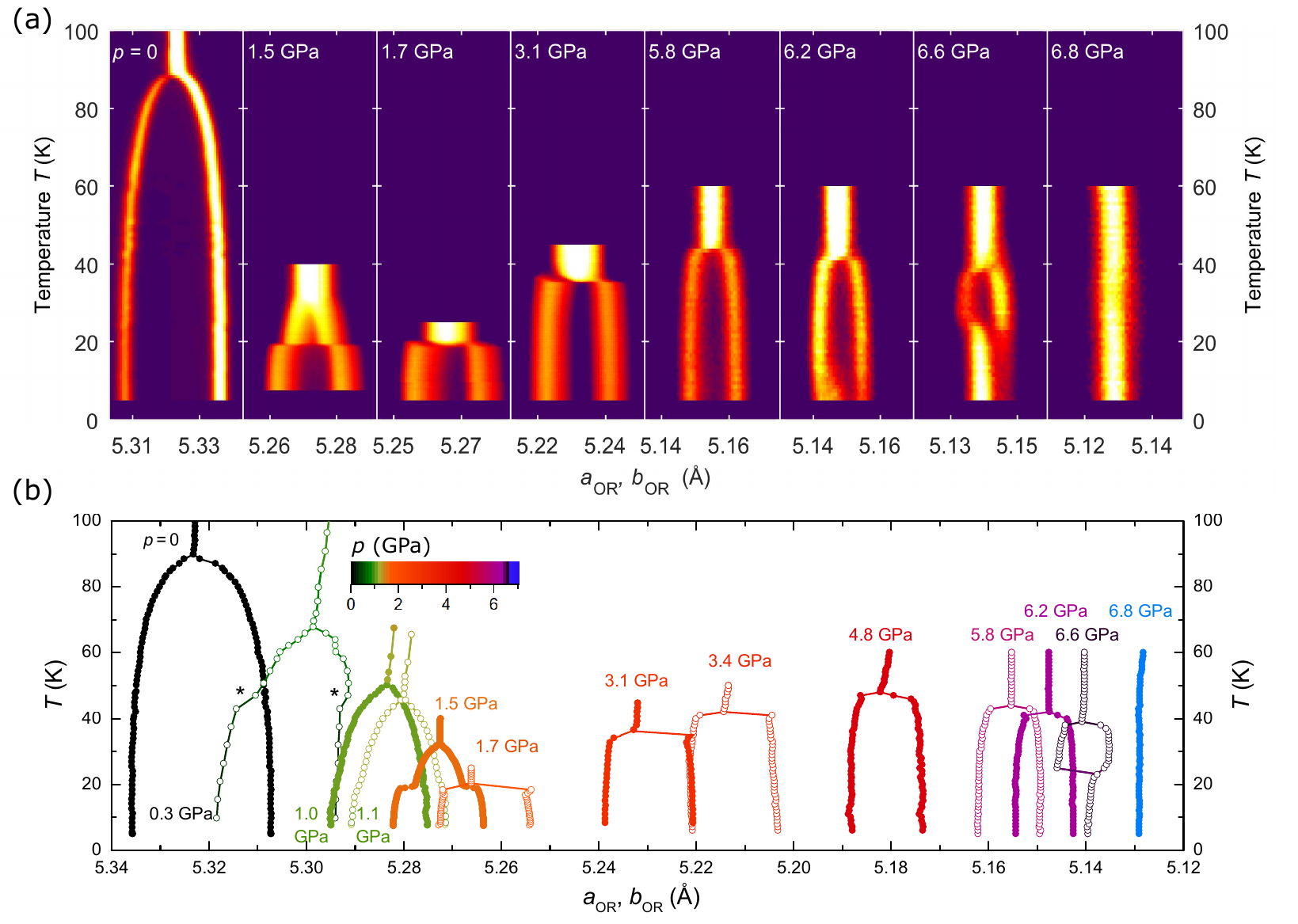}
	\caption{Temperature dependence of the in-plane orthorhombic lattice parameters $a_\mathrm{OR}$ and $b_\mathrm{OR}$ of FeSe for various pressures. (a) The detector intensity for positions spanning the $(3~3~0)$ or $(6~6~0)$ tetragonal unit cell Bragg peaks integrated over the transverse scattering directions at various pressures. Low-pressure ($p\leq3.1$ GPa) data are taken from Ref.\ \onlinecite{Kothapalli2016}. (b) Orthorhombic lattice parameters $a_\mathrm{OR}$ and $b_\mathrm{OR}$ as a function of temperature for various pressures. Stars mark the crossing of the He-solidification line, which entails a change in pressure (see color scale). Note that the horizontal scale decreases from left to right.}	
	\label{fig:1}
\end{figure*}

Figures \ref{fig:2b} and \ref{fig:1} show the results of high-energy x-ray diffraction measurements made close to $(H~H~0)$-type Bragg peaks and reveal the temperature and pressure dependence of the orthorhombic distortion. In Fig.\ \ref{fig:2b} data on samples from batches A (obtained with the Pixirad-$1$ detector) and B (obtained with the MAR$345$ detector) are compared. An obvious difference in peak profiles results from the different orders of the chosen Bragg peaks, the use of different detectors, and the broader mosaicity of samples from batch B. A low-temperature peak-splitting or broadening indicates an orthorhombic ground state. Structural transitions are clearly visible for samples from both batches. At the highest pressures of $6.8$ and $7.4$ GPa, the absence of any peak splitting or broadening and the temperature independent peak profiles up to $60$ K indicate a tetragonal ground state. 

As already shown in Ref.\ \onlinecite{Kothapalli2016}, the ambient-pressure second-order tetragonal-to-orthorhombic transition at $T_s$ is suppressed under pressure. At $p\gtrsim1.5$ GPa, a first-order transition occurs at $T_N<T_s$ and the two transitions merge for slightly higher pressures. This first-order transition is observed in the same manner up to $5.8$ GPa. To investigate the pressure-evolution of the orthorhombic distortion of FeSe in more detail, samples from batch A have been studied with additional fine pressure steps between 5.8 and 6.8 GPa (Figure \ref{fig:1}). 

At pressures above 5.8 GPa, a new behavior is observed, which is most pronounced at $p=6.6$ GPa. On decreasing temperature, the sample first undergoes the first-order tetragonal-to-orthorhombic transition, at $T_{s,N}=39$ K, before it transforms back into a tetragonal structure at $T_r=25$ K. At the just slightly higher pressure of 6.8 GPa, the sample remains tetragonal at all temperatures. We note that a small phase fraction ($\lesssim15\%$) appears to become orthorhombic in a limited temperature range even at 6.8 GPa, apparently experiencing a slightly lower effective pressure due to small internal stresses. Similarly, at the lower pressure of 6.2 GPa, a tetragonal fraction of the sample ($\sim20\%$) coexists with the major orthorhombic phase at base temperature. On heating, this phase fraction transforms to orthorhombic at $T_r\sim 12$~K. This "structurally reentrant" behavior is reminiscent of the "structural reentrance" in hole-doped 122-type materials upon transition into the tetragonal magnetic phase\cite{Avci2014,Boehmer2015,Allred2016,Taddei2017}.

Figure \ref{fig:1}(b) shows the in-plane lattice parameters of the majority phase of FeSe vs. temperature for all the studied pressures. FeSe has a high compressibility and the tetragonal in-plane lattice parameter is decreased by 3.7\% at $7$ GPa. In the pressure range 1.7-4.8 GPa, the orthorhombic transition results in an asymmetric change of in-plane lattice parameters so that the average of the $a$ and $b$ lattice constants decreases at the transition, similar to the Ba(Fe,Co)$_2$As$_2$ system\cite{Meingast2012}. This reverses at higher pressures, so that at $\sim6$ GPa, the $a$-$b$ average increases on cooling through $T_s$, similarly to underdoped (Ba,K)Fe$_2$As$_2$\cite{Boehmer2015II}.

% This behavior is reminiscent of the "structural reentrance" in hole-doped 122-type materials on the transition into the C4, tetragonal magnetic phase\cite{Avci2014,Boehmer2015,Taddei2017,Allred2016}.

\begin{figure*}
	\includegraphics[width=17.2cm]{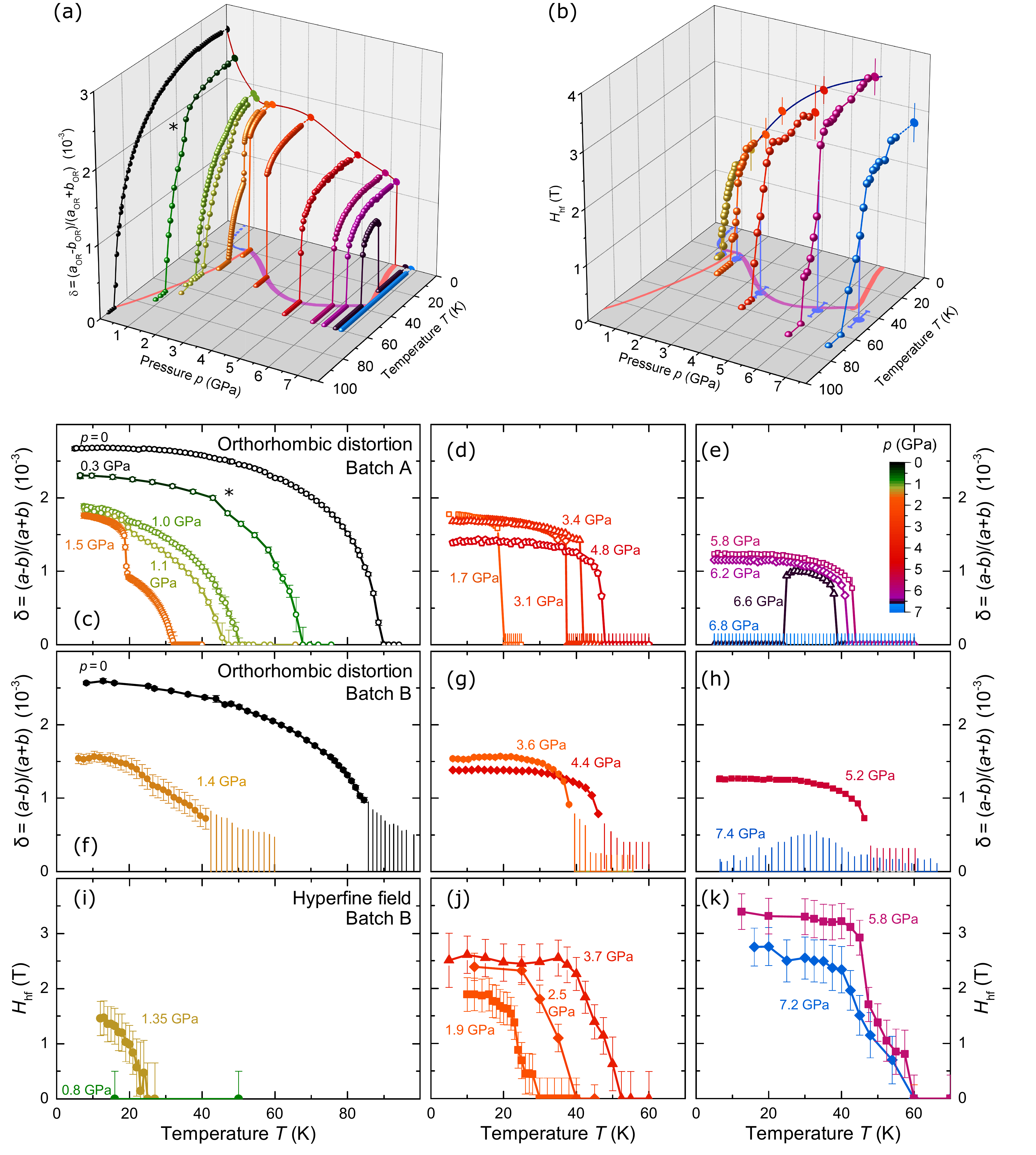}
	\caption{Structural and magnetic order parameters of FeSe under pressure. (a), (b) Three-dimensional representation of the temperature- and pressure-dependence of the orthorhombic distortion, $\delta=(a_\mathrm{OR}-b_\mathrm{OR})/(a_\mathrm{OR}+b_\mathrm{OR})$, and magnetic hyperfine field, $H_{\mathrm{hf}}$. Projections of the extrapolated low-temperature values are shown in the $T=0$ plane. Lines are guides to the eye. The phase diagram from Fig. \ref{fig:6} is indicated in the basal, $T-p$ plane. (c)-(e) Orthorhombic distortion $\delta$ of samples from batch A as a function of temperature at various color coded pressures [see scale in (e)]. The star symbol marks the crossing of the He solidification line. (f)-(h) Orthorhombic distortions $\delta$ of samples from batch B. Vertical bars represent a possible inhomogeneous distortion above the midpoint of the structural transition deduced from peak broadening. (i)-(k) Hyperfine field $H_\mathrm{hf}$ at the Fe site for samples from batch B, determined from the fitting shown in Fig. \ref{fig:2}. Data at 2.5 GPa from Ref. \onlinecite{Kothapalli2016} are also reported. Error bars represent the estimated total uncertainty including systematic errors.}	
	\label{fig:3}
\end{figure*}

\subsection{Structural and magnetic order parameters}

The temperature and pressure evolution of the orthorhombic order parameter $\delta=(a_\mathrm{OR}-b_\mathrm{OR})/(a_\mathrm{OR}+b_\mathrm{OR})$ and the magnetic hyperfine field $H_\mathrm{hf}$ at the iron site are presented in Fig. \ref{fig:3}. Here, we also compare the diffraction results for the two different batches. The transitions are very sharp in samples from batch A, whereas samples from batch B exhibit clear tails to the transitions, which likely arise from inhomogeneities caused by internal stresses. The hyperfine fields deduced from the NFS spectra in Fig. \ref{fig:3}(j,k) likely exhibit similar tails of the same origin. Nevertheless, transition temperatures can be well defined as the points of highest rate change of order parameters and the determined values of $T_s$ agree well between the two batches (see Fig. \ref{fig:6}). 

A second-order structural transition occurs in the absence of magnetic order at the lowest pressures $p\lesssim 0.8$ GPa. At the slightly higher pressure of 1.35 GPa, magnetic order emerges in a second-order like transition at $T_N<T_s$. In an intermediate pressure range $\sim1.7-6$ GPa, magnetic and structural transitions are both first order and are firmly coupled, as discussed in Ref. \onlinecite{Kothapalli2016}. 
 The coupling between structural and magnetic order surprisingly breaks down at pressures $p\gtrsim6.6$ GPa, when FeSe exhibits a tetragonal ground state even though an ordered magnetic hyperfine field is still observed at a pressure as high as 7.2 GPa. This indicates the presence of a tetragonal magnetic phase. 
 
We note that the tetragonal magnetic state in (Sr,Na)Fe$_2$As$_2$ was shown to be a coherent superposition of two spin-density waves and is characterized by two distinct Fe sites - one with zero and one with double the hyperfine field of the regular stripe-type phase\cite{Allred2016}. Unfortunately, our time-domain M\"ossbauer spectroscopic experiment is unable to distinguish such a state from a state of the same hyperfine field on all Fe sites. Here, we show the results of fitting with a single hyperfine field, $H_\mathrm{hf}$. A model in which zero moment is imposed for half of the Fe sites is able to fit our data at 7.2 GPa equally well, with a fitted value of $H_\mathrm{hf}\approx 4.5$ T for the moment-bearing Fe-sites at base temperature. Nevertheless, it is also possible that FeSe exhibits a completely different type of magnetic order, as might be indicated by the presence of N\'eel type magnetic fluctuations at ambient pressure\cite{Wang2015GS}.
 
Figures \ref{fig:3}(a)-(b) [see also Fig.\ \ref{fig:6}(b)] summarize the temperature and pressure evolution of the structural and magnetic order parameters over the phase diagram. The low-temperature value of the magnetic hyperfine field of the orthorhombic-magnetic phase increases monotonically under pressure up to $\sim6$ GPa, even though the magnetic transition temperature has a dome-like pressure dependence and peaks around 4.8 GPa. 
 %The fitted $H_\mathrm{hf}$ is somewhat lower in the C4,  tetragonal magnetic state at high pressures. 
The orthorhombic distortion exhibits a complex temperature-pressure dependence. Initially, on increasing pressure, the value of the orthorhombic distortion is suppressed. The low-temperature value of $\delta$ barely changes as a function of pressure between $1-3.4$ GPa and then gradually decreases under pressure over the range at which $T_\mathrm{s,N}$ has a maximum. The structurally "reentrant" tetragonal behavior with vanishing lattice distortion at the lowest temperatures is limited over a very small pressure range $6.2\lesssim p<6.8$ GPa and the orthorhombic distortion is absent at all temperatures at 6.8 GPa. 
		
\section{Comparison of F\MakeLowercase{e}S\MakeLowercase{e} and B\MakeLowercase{a}F\MakeLowercase{e}$_2$A\MakeLowercase{s}$_2$-based superconductors}

\begin{figure}
	\includegraphics[width=8.6cm]{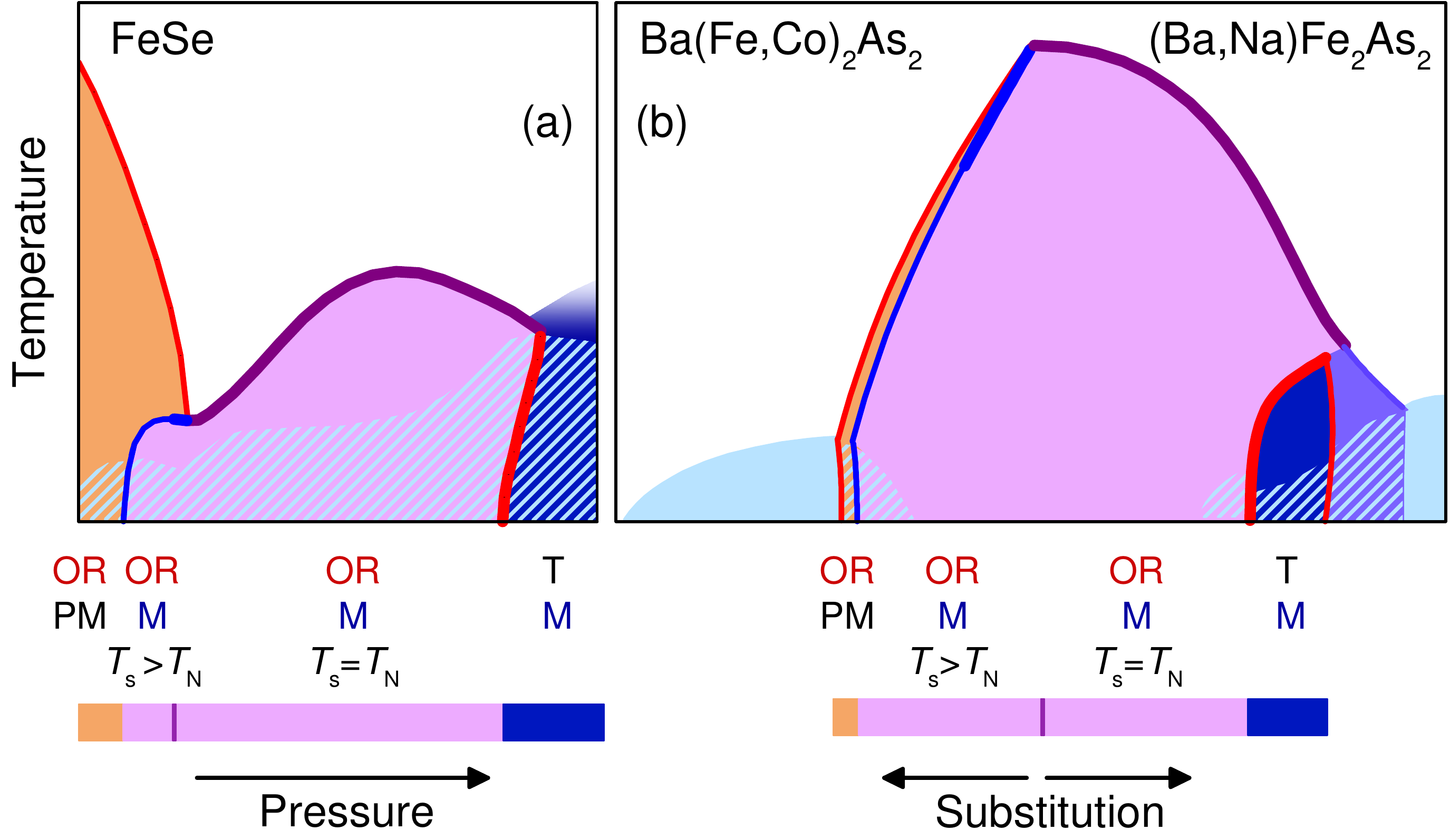}
	\caption{Schematic phase diagrams of FeSe under pressure (this work) and BaFe$_2$As$_2$ doped with cobalt (Ref. \onlinecite{Hardy2010}) and sodium (Ref. \onlinecite{Wang2017II}). The phase lines and phases are color-coded as in Fig. \ref{fig:6}. The evolution of magnetic/structural ground states from orthorhombic/paramagnetic (OR+PM), via orthorhombic/magnetically ordered (OR+M) with $T_s>T_N$, orthorhombic/magnetically ordered with $T_s=T_N$ to tetragonal/magnetically ordered (T+M) for increasing pressure in FeSe parallels the evolution of the ground states of BaFe$_2$As$_2$ from electron-doped to hole-doped.}	
	\label{fig:7}
\end{figure}

The schematic temperature-pressure phase diagram of FeSe in Fig. \ref{fig:7}(a) has remarkable similarities to the temperature-substitution phase diagram of the canonical BaFe$_2$As$_2$ iron-based superconductors [\ref{fig:7}(a)]. 
%In these 122-type systems, the role of substitutions has been intensively discussed\cite{Wadati2010}, and the respective roles of charge-doping, disorder and chemical pressure are still controversial\cite{Merz2016,Herbig2016,others}. 
In particular, we consider substitution of Co on the Fe-site ("electron-doping") and substitution of Na on the Ba site ("hole-doping") as a way of tuning BaFe$_2$As$_2$ quasi-continuously as shown in Figure \ref{fig:7}(b). Then, the sequence of magnetic/structural ground states and the topology of the phase diagrams formed by the corresponding phase lines are similar. FeSe at low pressures and slightly underdoped Ba(Fe,Co)$_2$As$_2$ both have an orthorhombic paramagnetic ground state. On increasing pressure or decreasing Co content the ground state changes to orthorhombic and antiferromagnetic within a region for which $T_s>T_N$ until the structural and magnetic phase lines merge. On the high-pressure side of FeSe and in close to optimally doped (Ba,Na)Fe$_2$As$_2$ a tetragonal magnetic ground state emerges. 

\begin{figure}
	\includegraphics[width=8.6cm]{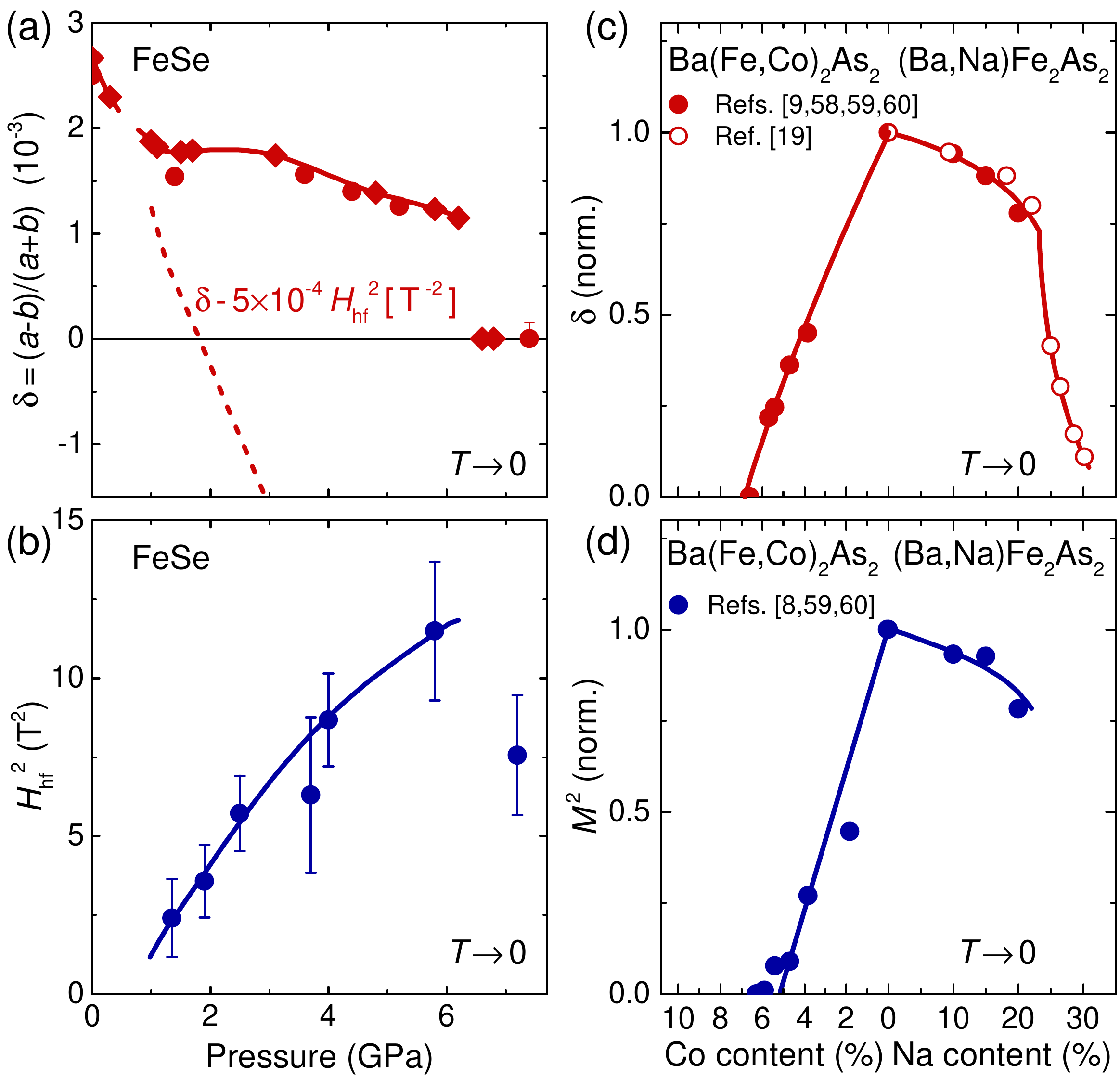}
	\caption{Evolution of the orthorhombic and magnetic order parameters of FeSe and BaFe$_2$As$_2$. (a) Low-temperature extrapolation of $\delta$ of FeSe as a function of pressure. The dashed line shows the experimental $\delta$ subtracted by a contribution proportional to $H_{\mathrm {hf}}^2$, which continues the trend of $\delta(p)$ from low pressures (in the absence of magnetic order). (b) Square of hyperfine field $H_{\mathrm {hf}}^2$ of FeSe under pressure. (c) Low-temperature extrapolation of $\delta$ of Ba(Fe,Co)$_2$As$_2$ (Refs. \onlinecite{Nandi2010,Prozorov2009}) and (Ba,Na)Fe$_2$As$_2$ from Refs. \onlinecite{Avci2012,Avci2013} (full symbols) and \onlinecite{Wang2017II} (open symbols). (d) Square of the ordered magnetic moment of Ba(Fe,Co)$_2$As$_2$ \cite{Fernandes2010} and (Ba,Na)Fe$_2$As$_2$ \cite{Avci2012,Avci2013}. Lines are a guide to the eye. The data from different references in (c),(d) are scaled at pure BaFe$_2$As$_2$.  Whereas the order parameters (approximately) obey linear-quadratic coupling $\delta \propto M^2$ in the 122-type system, this is not the case for FeSe.}
	\label{fig:8}
\end{figure}

%are considered and Co or Ni substitution on the Fe-site ("electron-doping") are common. Whereas both kinds of substitution result in similar phase diagrams, the differences have emerged more and more clearly over the years. The splitting of $T_s$ and $T_N$ and the ensuing presence of purely nematic phase occurs only upon electron doping. Upon hole-doping, the structural transition remains tied to the magnetic transition. However, C4, tetragonal magnetic phases are prone to emerge upon hole doping [Refs]. Continuous tuning between these two extreme regimes of a purely nematic phase and a C4, tetragonal magnetic phase has so far remained elusive. Here, we establish that FeSe can be tuned between those two regimes continuously using pressure as a well-defined tuning parameter.
%This brings the iron-chalcogenide FeSe "back into the family" of iron-arsenides. Note that deMedici et al.\cite{deMedici2014} proposed that (Ba,K)Fe2As2 and Ba(Fe,Co)2As2 form a single superconducting dome, tuned by the filling of the iron $d$ orbitals, that is only accidentally "suppressed" by a dome of stripe-type magnetic order. 

A qualitative difference between FeSe under pressure and substituted BaFe$_2$As$_2$ becomes evident when the value of the orthorhombic and magnetic order parameters are considered. In the 122-type systems, the orthorhombic distortion, $\delta$, and the ordered magnetic moment, $M$, in the low-temperature limit follow each other closely. Linear-quadratic coupling $\delta\propto M^2$ is theoretically expected\cite{Barzykin2009,Qi2009,Fernandes2010,Cano2010}. It has, for example, explicitly been shown experimentally for (Ba,K)Fe$_2$As$_2$ \cite{Avci2012}, which features strongly coupled magneto-structural transitions. This proportionality does not hold for FeSe, where $H_\mathrm{hf}\propto M$ increases monotonically up to $\sim7$ GPa and $\delta$ has a complex non-monotonic pressure dependence (Figure \ref{fig:8}).
%Could the refernces for linear-quadratic relation be Avci et al.,PRB 85,184507(2012) and Wilson et al, PRB 79,185419(2009) - Avci: yes, I added that. Wilson says that is should be linear-quadratic, but experimentally he finds quadratic-quadratic: quite puzzling to me. 

One way to rationalize our results on FeSe is to assume that the material's tendency for an orthorhombic distortion at ambient pressure is independent of the pressure-induced magnetic order. Under increasing pressure, this structural instability is weakened and $\delta$ decreases, whereas the magnetic instability is strengthened. The low cost of orthorhombic distortion at moderate pressures likely favors a stripe-type antiferromagnetic order. %Such a magnetic ground state has indeed been proposed on the basis of several experiments\cite{muSR, NMR}, though its definite proof is still lacking. 
By symmetry, an orthorhombic distortion and stripe-type magnetic order couple cooperatively so that the low-temperature value of $\delta$ should increase by an amount $\propto M^2$, whenever both types of order occur. If such a contribution to the experimental value of $\delta$ is subtracted, the initial trend of $\delta(p)$ indeed continues smoothly [dashed line in Fig. \ref{fig:8}(b)]. Symmetry also dictates that structural and magnetic phase lines merge when $T_s$ meets $T_N$, because the stripe-type AFM state is necessarily orthorhombic. Only above $\sim6.6 $ GPa does any orthorhombic distortion become so unfavorable that tetragonal magnetic order is established instead of stripe-type antiferromagnetism. Thus, the origin of the orthorhombic distortion of FeSe at ambient pressure is likely independent from the pressure-induced magnetic order. Theoretically, the nematic order of FeSe at ambient pressure has been found to arise from a Pomeranchuk instability within a renormalization group analysis\cite{Chubukov2016}, which could explain the observed result. Similarly, the theoretically proposed antiferroquadrupolar order\cite{Yu2015} is a candidate. The effect of pressure as a tuning parameter has been subject to numerous theoretical studies as well\cite{Glasbrenner2015,Yamakawa_PRX_16,Scherer_FeSe,Ishizuka2017}. Model\cite{Scherer_FeSe} and ab-initio\cite{Ishizuka2017} calculations find a decrease of the tendency to charge order under pressure that could be associated with our results.

%For an itinerant system, the value of the ordered moment does not have to be the same for different types of antiferromagnetic order.

%A delicate energy balance is likely the cause of the "structural reentrance" at 6.6 GPa. 

%In the 2-6 GPa range, the magnetic order gets stronger as shown by the increased value of $H_\mathrm{hf}$. However, the increasing cost of orthorhombic distortion and the coupling of the two types of order leads to a dome-like pressure dependence of $T_\mathrm{s,N}$.  We speculate that the orthorhombic magnetic order in FeSe has the same stripe-type magnetic wave vector as other iron-based materials, as supported by similar Fermi surfaces of the two material classes and the observed spin fluctuations related to the stripe-type wavevectors. If the type of ordering wavevector is not changing, the C4, tetragonal magnetic state in FeSe is similarly likely a superposition of two stripe-type spin-density waves. 
%The coupling is there is both systems, however, for FeSe, there is a distinct origin of the nematic phase and magnetic order. It just "accidentally" couple. This explains the topology the of the phase diagram and the nonmonotonic pressure dependence of orthorhombic distortion. 

The similarity between FeSe under pressure and substituted BaFe$_2$As$_2$ does not hold for the pressure dependence of superconductivity. In Fig. \ref{fig:6} we report $T_c$ as determined from zero resistance\cite{Sun_2016} and the onset of diamagnetism\cite{Miyoshi2014}. However, it has been questioned whether superconductivity can microscopically coexists with magnetic order in FeSe. The dc magnetization in Ref. \onlinecite{Miyoshi2014} indicates an unchanged amount of diamagnetic shielding up to 7 GPa. Nevertheless, the superconducting transition becomes significantly broader in the pressure range in which orthorhombic magnetism and superconductivity overlap\cite{Terashima2015,Sun_2016} and the spin-lattice relaxation rate in NMR does not change at the putative $T_c$\cite{Wang2016}, questioning bulk superconductivity. Similarly, such a coexistence has been questioned in Fe(Se,S)\cite{Yip2017}. None of our diffraction or NFS results show any anomaly that could be associated with $T_c$. Since the question of bulk superconductivity in FeSe under pressure is not solved, any conclusion about relation between the tetragonal magnetic phase and superconductivity at high pressures has to be speculative. From the literature, it seems that $T_c$ is maximized in this pressure range. We note that in hole-doped 122's, $T_c$ tends to be slighlty suppressed\cite{Boehmer2015II,Wang2017II} by the presence of the tetragonal magnetic phase.
%Although it is for sulphur doped FeSe, could we refer Matsuure(2017) to support the doubts about bulk supercondctivity? - sure, but Yip(2017) is more explicit - done. 

\section{Conclusion}
In summary, we have studied the structural and magnetic phase diagram and order parameters of FeSe. We have exposed an unexpected variability of the pressure-induced magnetic order. It couples cooperatively with an orthorhombic distortion in the intermediate pressure range. However, at higher pressures, we have discovered the complete suppression of orthorhombic distortion and a tetragonal magnetic phase. This sets the stage for high-temperature superconductivity in FeSe under pressure. 
The topology of the temperature-pressure phase diagram of FeSe resembles closely the well-known phase diagram of electron/hole-doped BaFe$_2$As$_2$. The origin of the nematic phase in BaFe$_2$As$_2$-based materials is widely accepted to be tied to their stripe-type antiferromagnetism. Here, the pressure evolution of the magnetic and structural order parameters of FeSe leads us to suggest that the origin of the orthorhombic phase in FeSe is distinct from the pressure induced magnetic order. The cooperative coupling between orthorhombicity and magnetic order that is present in both FeSe and BaFe$_2$As$_2$-based systems likely leads to the similarities in the phase diagrams.

\section{Acknowledgements}
We would like to acknowledge the assistance of D. S. Robinson, C. Benson, S. Tkachev, S. G. Sinogeikin, Ross Hrubiak, Barbara Lavina, R. Somayazulu and M. Baldini, and helpful discussions with R. J. McQueeney. Work at the Ames Laboratory was supported by the Department of Energy, Basic Energy Sciences, Division of Materials Sciences \& Engineering, under Contract No. DE-AC02-07CH11358.  This research used resources of the Advanced Photon Source, a U.S. Department of Energy (DOE) Office of Science User Facility operated for the DOE Office of Science by Argonne National Laboratory under Contract No. DE-AC02-06CH11357. HPCAT operations are supported by DOE-NNSA under Award No. DE-NA0001974, with partial instrumentation funding by NSF. Use of the COMPRES-GSECARS gas loading system was supported by COMPRES under NSF Cooperative Agreement EAR 11-57758 and by GSECARS through NSF grant EAR-1128799 and DOE grant DE-FG02-94ER14466. Y.X. acknowledges the support of DOE-BES/DMSE under Award DE-FG02-99ER45775. W.B. acknowledges the partial support by COMPRES, the Consortium for Materials Properties Research in Earth Sciences under NSF Cooperative Agreement EAR 1606856. 
        
\appendix
\section{High-pressure phase of FeSe}

\begin{figure}
	\includegraphics[width=8.6cm]{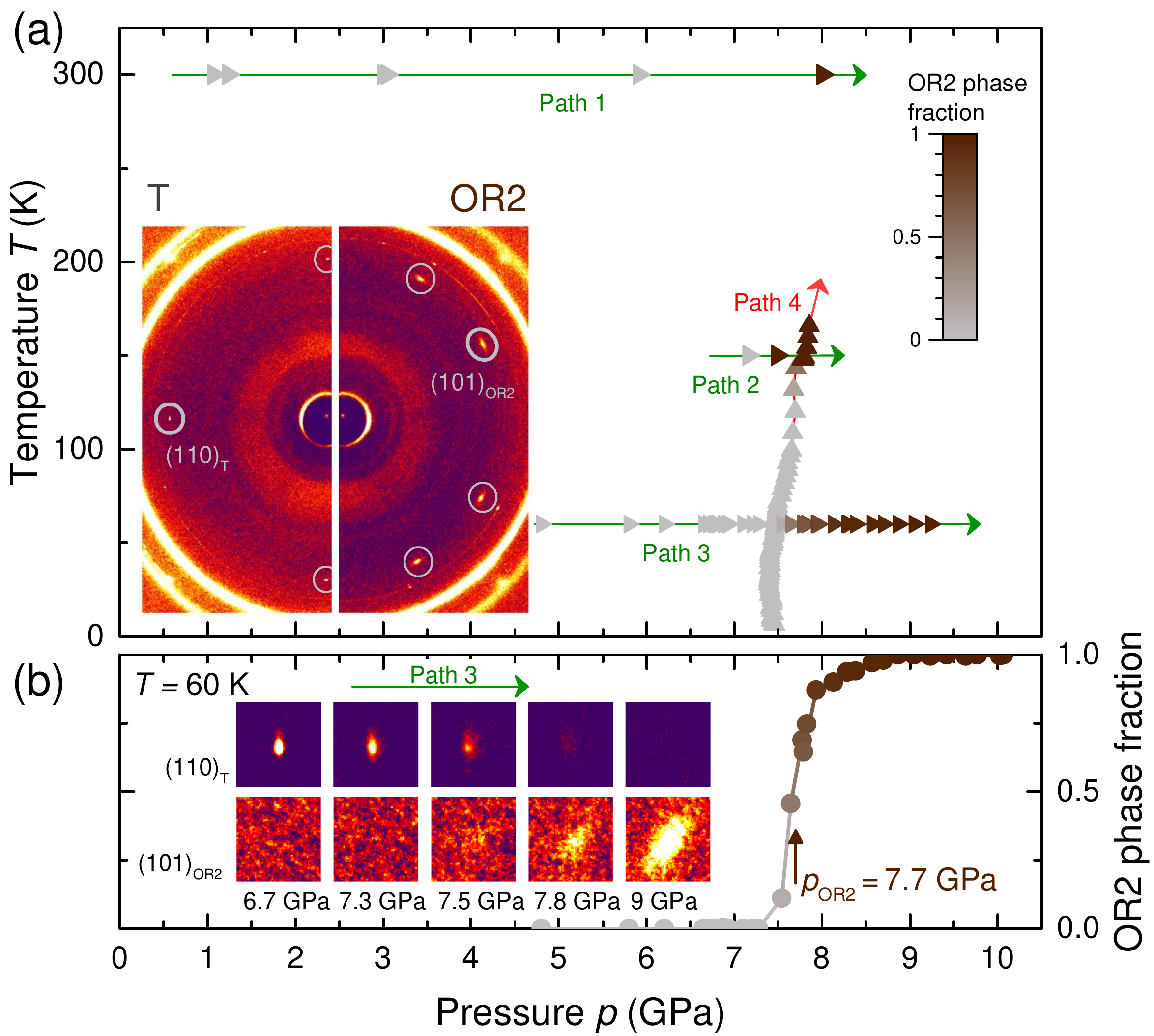}
	\caption{The high-pressure structural transition of FeSe. (a) Points in the temperature-pressure phase diagram of FeSe with the fraction of the OR2 phase color-coded. The transition into the OR2 phase was observed to be abrupt and irreversible. The inset shows sections of the ($HK$0) diffraction plane in the T and OR2 phases. The $(110)_\mathrm{T}$ and $(101)_\mathrm{OR2}$ and symmetry-equivalent reflections are highlighted by gray circles. (b) Phase fraction of the OR2 phase as evaluated by the relative intensity of the $(101)_\mathrm{OR2}$ with respect to the $(110)_\mathrm{T}$) Bragg reflections on increasing pressure at constant $T=60$ K (path 3). The insets show the area surrounding these Bragg reflections at representative pressures.}	
	\label{fig:5}
\end{figure}

At just slightly higher pressures than those presented in Fig. \ref{fig:3} the samples of both batches undergo a severe structural transition illustrated in Fig. \ref{fig:5}. This transition is well-known in literature\cite{Medvedev2009,Margadonna2009,Kumar2010,Svitlyk2016,Lebert2017} and the high-pressure orthorhombic "OR2" phase has been identified as having MnP-type structure with a dramatic volume reduction of $\sim10\%$ with respect to the layered PbO-type FeSe-phase that is stable at lower pressures. The inset in Fig. \ref{fig:5}(a) shows sections of the ($HK0$) scattering plane of the two phases at representative pressures. The sharp $(110)_\mathrm{T}$ Bragg peaks of the tetragonal phase completely disappear in the high-pressure OR2 phase. The latter is characterized by 8 much broader $(101)_\mathrm{OR2}$ type reflections, indicating significant sample degradation which is clearly due to the huge volume reduction and build-up of stress due to domains of the MnP-type structure with different orientations. We have observed this transition in measurements with increasing pressure at constant temperatures 60 K, 150 K and 300 K, in three different samples (from batch A) and also on temperature increase (concomitant with a slight pressure increase, see path 4 in the Fig. \ref{fig:5}) in a sample from batch B. The fine pressure steps along path 3 allow to resolve particularly well the rapidly changing OR2 phase fraction between $p=7.5-8.5$ GPa. Together, these four independent measurements define a sharp and nearly temperature-independent phase line at $p\approx7.7$ GPa. 
We found this transition to have a severe hysteresis and to cause irreversible changes to the single-crystalline sample. For example, no return to the tetragonal phase was observed even after decreasing pressure down to 3 GPa and increasing temperature to 300 K following the measurement along path 2.

In previous reports using polycrystalline material\cite{Medvedev2009,Margadonna2009,Kumar2010} and in a single-crystal study with glycerol pressure medium\cite{Sun_2016}, the structural transition into the OR2 phase has been observed at $p\sim10-12$ GPa and often with a significant phase coexistence range. Our lower critical pressure is, however, close to the one reported in Refs. \onlinecite{Svitlyk2016,Lebert2017}, which also used single crystals and He as pressure medium. This transition into a much more closely-packed crystal structure at high pressures marks the end of the stability of the layered structure of FeSe that shares its structural motive with iron-arsenide superconductors. 

\bibliographystyle{apsrev4-1}
\bibliography{referencespnictides}

\begin{thebibliography}{69}%
\makeatletter
\providecommand \@ifxundefined [1]{%
 \@ifx{#1\undefined}
}%
\providecommand \@ifnum [1]{%
 \ifnum #1\expandafter \@firstoftwo
 \else \expandafter \@secondoftwo
 \fi
}%
\providecommand \@ifx [1]{%
 \ifx #1\expandafter \@firstoftwo
 \else \expandafter \@secondoftwo
 \fi
}%
\providecommand \natexlab [1]{#1}%
\providecommand \enquote  [1]{``#1''}%
\providecommand \bibnamefont  [1]{#1}%
\providecommand \bibfnamefont [1]{#1}%
\providecommand \citenamefont [1]{#1}%
\providecommand \href@noop [0]{\@secondoftwo}%
\providecommand \href [0]{\begingroup \@sanitize@url \@href}%
\providecommand \@href[1]{\@@startlink{#1}\@@href}%
\providecommand \@@href[1]{\endgroup#1\@@endlink}%
\providecommand \@sanitize@url [0]{\catcode `\\12\catcode `\$12\catcode
  `\&12\catcode `\#12\catcode `\^12\catcode `\_12\catcode `\%12\relax}%
\providecommand \@@startlink[1]{}%
\providecommand \@@endlink[0]{}%
\providecommand \url  [0]{\begingroup\@sanitize@url \@url }%
\providecommand \@url [1]{\endgroup\@href {#1}{\urlprefix }}%
\providecommand \urlprefix  [0]{URL }%
\providecommand \Eprint [0]{\href }%
\providecommand \doibase [0]{http://dx.doi.org/}%
\providecommand \selectlanguage [0]{\@gobble}%
\providecommand \bibinfo  [0]{\@secondoftwo}%
\providecommand \bibfield  [0]{\@secondoftwo}%
\providecommand \translation [1]{[#1]}%
\providecommand \BibitemOpen [0]{}%
\providecommand \bibitemStop [0]{}%
\providecommand \bibitemNoStop [0]{.\EOS\space}%
\providecommand \EOS [0]{\spacefactor3000\relax}%
\providecommand \BibitemShut  [1]{\csname bibitem#1\endcsname}%
\let\auto@bib@innerbib\@empty
%</preamble>
\bibitem [{\citenamefont {Dai}(2015)}]{Dai15_review}%
  \BibitemOpen
  \bibfield  {author} {\bibinfo {author} {\bibfnamefont {P.}~\bibnamefont
  {Dai}},\ }\href {\doibase 10.1103/RevModPhys.87.855} {\bibfield  {journal}
  {\bibinfo  {journal} {Rev. Mod. Phys.}\ }\textbf {\bibinfo {volume} {87}},\
  \bibinfo {pages} {855} (\bibinfo {year} {2015})}\BibitemShut {NoStop}%
\bibitem [{\citenamefont {Cano}\ \emph {et~al.}(2010)\citenamefont {Cano},
  \citenamefont {Civelli}, \citenamefont {Eremin},\ and\ \citenamefont
  {Paul}}]{Cano2010}%
  \BibitemOpen
  \bibfield  {author} {\bibinfo {author} {\bibfnamefont {A.}~\bibnamefont
  {Cano}}, \bibinfo {author} {\bibfnamefont {M.}~\bibnamefont {Civelli}},
  \bibinfo {author} {\bibfnamefont {I.}~\bibnamefont {Eremin}}, \ and\ \bibinfo
  {author} {\bibfnamefont {I.}~\bibnamefont {Paul}},\ }\href {\doibase
  10.1103/PhysRevB.82.020408} {\bibfield  {journal} {\bibinfo  {journal} {Phys.
  Rev. B}\ }\textbf {\bibinfo {volume} {82}},\ \bibinfo {pages} {020408}
  (\bibinfo {year} {2010})}\BibitemShut {NoStop}%
\bibitem [{\citenamefont {Sapkota}\ \emph {et~al.}(2014)\citenamefont
  {Sapkota}, \citenamefont {Tucker}, \citenamefont {Ramazanoglu}, \citenamefont
  {Tian}, \citenamefont {Ni}, \citenamefont {Cava}, \citenamefont {McQueeney},
  \citenamefont {Goldman},\ and\ \citenamefont {Kreyssig}}]{Sapkota2014}%
  \BibitemOpen
  \bibfield  {author} {\bibinfo {author} {\bibfnamefont {A.}~\bibnamefont
  {Sapkota}}, \bibinfo {author} {\bibfnamefont {G.~S.}\ \bibnamefont {Tucker}},
  \bibinfo {author} {\bibfnamefont {M.}~\bibnamefont {Ramazanoglu}}, \bibinfo
  {author} {\bibfnamefont {W.}~\bibnamefont {Tian}}, \bibinfo {author}
  {\bibfnamefont {N.}~\bibnamefont {Ni}}, \bibinfo {author} {\bibfnamefont
  {R.~J.}\ \bibnamefont {Cava}}, \bibinfo {author} {\bibfnamefont {R.~J.}\
  \bibnamefont {McQueeney}}, \bibinfo {author} {\bibfnamefont {A.~I.}\
  \bibnamefont {Goldman}}, \ and\ \bibinfo {author} {\bibfnamefont
  {A.}~\bibnamefont {Kreyssig}},\ }\href {\doibase 10.1103/PhysRevB.90.100504}
  {\bibfield  {journal} {\bibinfo  {journal} {Phys. Rev. B}\ }\textbf {\bibinfo
  {volume} {90}},\ \bibinfo {pages} {100504} (\bibinfo {year}
  {2014})}\BibitemShut {NoStop}%
\bibitem [{\citenamefont {de~la Cruz}\ \emph {et~al.}(2008)\citenamefont {de~la
  Cruz}, \citenamefont {Huang}, \citenamefont {Lynn}, \citenamefont {Li},
  \citenamefont {II}, \citenamefont {Zarestky}, \citenamefont {Mook},
  \citenamefont {Chen}, \citenamefont {Luo}, \citenamefont {Wang},\ and\
  \citenamefont {Dai}}]{Cruz2008}%
  \BibitemOpen
  \bibfield  {author} {\bibinfo {author} {\bibfnamefont {C.}~\bibnamefont
  {de~la Cruz}}, \bibinfo {author} {\bibfnamefont {Q.}~\bibnamefont {Huang}},
  \bibinfo {author} {\bibfnamefont {J.~W.}\ \bibnamefont {Lynn}}, \bibinfo
  {author} {\bibfnamefont {J.}~\bibnamefont {Li}}, \bibinfo {author}
  {\bibfnamefont {W.~R.}\ \bibnamefont {II}}, \bibinfo {author} {\bibfnamefont
  {J.~L.}\ \bibnamefont {Zarestky}}, \bibinfo {author} {\bibfnamefont {H.~A.}\
  \bibnamefont {Mook}}, \bibinfo {author} {\bibfnamefont {G.~F.}\ \bibnamefont
  {Chen}}, \bibinfo {author} {\bibfnamefont {J.~L.}\ \bibnamefont {Luo}},
  \bibinfo {author} {\bibfnamefont {N.~L.}\ \bibnamefont {Wang}}, \ and\
  \bibinfo {author} {\bibfnamefont {P.}~\bibnamefont {Dai}},\ }\href {\doibase
  10.1038/nature07057} {\bibfield  {journal} {\bibinfo  {journal} {Nature}\
  }\textbf {\bibinfo {volume} {453}},\ \bibinfo {pages} {899} (\bibinfo {year}
  {2008})}\BibitemShut {NoStop}%
\bibitem [{\citenamefont {Ni}\ \emph {et~al.}(2008)\citenamefont {Ni},
  \citenamefont {Tillman}, \citenamefont {Yan}, \citenamefont {Kracher},
  \citenamefont {Hannahs}, \citenamefont {Bud'ko},\ and\ \citenamefont
  {Canfield}}]{Ni2008}%
  \BibitemOpen
  \bibfield  {author} {\bibinfo {author} {\bibfnamefont {N.}~\bibnamefont
  {Ni}}, \bibinfo {author} {\bibfnamefont {M.~E.}\ \bibnamefont {Tillman}},
  \bibinfo {author} {\bibfnamefont {J.-Q.}\ \bibnamefont {Yan}}, \bibinfo
  {author} {\bibfnamefont {A.}~\bibnamefont {Kracher}}, \bibinfo {author}
  {\bibfnamefont {S.~T.}\ \bibnamefont {Hannahs}}, \bibinfo {author}
  {\bibfnamefont {S.~L.}\ \bibnamefont {Bud'ko}}, \ and\ \bibinfo {author}
  {\bibfnamefont {P.~C.}\ \bibnamefont {Canfield}},\ }\href {\doibase
  10.1103/PhysRevB.78.214515} {\bibfield  {journal} {\bibinfo  {journal} {Phys.
  Rev. B}\ }\textbf {\bibinfo {volume} {78}},\ \bibinfo {pages} {214515}
  (\bibinfo {year} {2008})}\BibitemShut {NoStop}%
\bibitem [{\citenamefont {Chu}\ \emph {et~al.}(2009)\citenamefont {Chu},
  \citenamefont {Analytis}, \citenamefont {Kucharczyk},\ and\ \citenamefont
  {Fisher}}]{Chu2009}%
  \BibitemOpen
  \bibfield  {author} {\bibinfo {author} {\bibfnamefont {J.-H.}\ \bibnamefont
  {Chu}}, \bibinfo {author} {\bibfnamefont {J.~G.}\ \bibnamefont {Analytis}},
  \bibinfo {author} {\bibfnamefont {C.}~\bibnamefont {Kucharczyk}}, \ and\
  \bibinfo {author} {\bibfnamefont {I.~R.}\ \bibnamefont {Fisher}},\ }\href
  {\doibase 10.1103/PhysRevB.79.014506} {\bibfield  {journal} {\bibinfo
  {journal} {Phys. Rev. B}\ }\textbf {\bibinfo {volume} {79}},\ \bibinfo
  {pages} {014506} (\bibinfo {year} {2009})}\BibitemShut {NoStop}%
\bibitem [{\citenamefont {Fernandes}\ \emph {et~al.}(2014)\citenamefont
  {Fernandes}, \citenamefont {Chubukov},\ and\ \citenamefont
  {Schmalian}}]{Fernandes2014}%
  \BibitemOpen
  \bibfield  {author} {\bibinfo {author} {\bibfnamefont {R.~M.}\ \bibnamefont
  {Fernandes}}, \bibinfo {author} {\bibfnamefont {A.~V.}\ \bibnamefont
  {Chubukov}}, \ and\ \bibinfo {author} {\bibfnamefont {J.}~\bibnamefont
  {Schmalian}},\ }\href {\doibase 10.1038/nphys2877} {\bibfield  {journal}
  {\bibinfo  {journal} {Nature Physics}\ }\textbf {\bibinfo {volume} {10}},\
  \bibinfo {pages} {97} (\bibinfo {year} {2014})}\BibitemShut {NoStop}%
\bibitem [{\citenamefont {Fernandes}\ \emph {et~al.}(2010)\citenamefont
  {Fernandes}, \citenamefont {VanBebber}, \citenamefont {Bhattacharya},
  \citenamefont {Chandra}, \citenamefont {Keppens}, \citenamefont {Mandrus},
  \citenamefont {McGuire}, \citenamefont {Sales}, \citenamefont {Sefat},\ and\
  \citenamefont {Schmalian}}]{Fernandes2010}%
  \BibitemOpen
  \bibfield  {author} {\bibinfo {author} {\bibfnamefont {R.~M.}\ \bibnamefont
  {Fernandes}}, \bibinfo {author} {\bibfnamefont {L.~H.}\ \bibnamefont
  {VanBebber}}, \bibinfo {author} {\bibfnamefont {S.}~\bibnamefont
  {Bhattacharya}}, \bibinfo {author} {\bibfnamefont {P.}~\bibnamefont
  {Chandra}}, \bibinfo {author} {\bibfnamefont {V.}~\bibnamefont {Keppens}},
  \bibinfo {author} {\bibfnamefont {D.}~\bibnamefont {Mandrus}}, \bibinfo
  {author} {\bibfnamefont {M.~A.}\ \bibnamefont {McGuire}}, \bibinfo {author}
  {\bibfnamefont {B.~C.}\ \bibnamefont {Sales}}, \bibinfo {author}
  {\bibfnamefont {A.~S.}\ \bibnamefont {Sefat}}, \ and\ \bibinfo {author}
  {\bibfnamefont {J.}~\bibnamefont {Schmalian}},\ }\href {\doibase
  10.1103/PhysRevLett.105.157003} {\bibfield  {journal} {\bibinfo  {journal}
  {Phys. Rev. Lett.}\ }\textbf {\bibinfo {volume} {105}},\ \bibinfo {pages}
  {157003} (\bibinfo {year} {2010})}\BibitemShut {NoStop}%
\bibitem [{\citenamefont {Nandi}\ \emph {et~al.}(2010)\citenamefont {Nandi},
  \citenamefont {Kim}, \citenamefont {Kreyssig}, \citenamefont {Fernandes},
  \citenamefont {Pratt}, \citenamefont {Thaler}, \citenamefont {Ni},
  \citenamefont {Bud'ko}, \citenamefont {Canfield}, \citenamefont {Schmalian},
  \citenamefont {McQueeney},\ and\ \citenamefont {Goldman}}]{Nandi2010}%
  \BibitemOpen
  \bibfield  {author} {\bibinfo {author} {\bibfnamefont {S.}~\bibnamefont
  {Nandi}}, \bibinfo {author} {\bibfnamefont {M.~G.}\ \bibnamefont {Kim}},
  \bibinfo {author} {\bibfnamefont {A.}~\bibnamefont {Kreyssig}}, \bibinfo
  {author} {\bibfnamefont {R.~M.}\ \bibnamefont {Fernandes}}, \bibinfo {author}
  {\bibfnamefont {D.~K.}\ \bibnamefont {Pratt}}, \bibinfo {author}
  {\bibfnamefont {A.}~\bibnamefont {Thaler}}, \bibinfo {author} {\bibfnamefont
  {N.}~\bibnamefont {Ni}}, \bibinfo {author} {\bibfnamefont {S.~L.}\
  \bibnamefont {Bud'ko}}, \bibinfo {author} {\bibfnamefont {P.~C.}\
  \bibnamefont {Canfield}}, \bibinfo {author} {\bibfnamefont {J.}~\bibnamefont
  {Schmalian}}, \bibinfo {author} {\bibfnamefont {R.~J.}\ \bibnamefont
  {McQueeney}}, \ and\ \bibinfo {author} {\bibfnamefont {A.~I.}\ \bibnamefont
  {Goldman}},\ }\href {\doibase 10.1103/PhysRevLett.104.057006} {\bibfield
  {journal} {\bibinfo  {journal} {Phys. Rev. Lett.}\ }\textbf {\bibinfo
  {volume} {104}},\ \bibinfo {pages} {057006} (\bibinfo {year}
  {2010})}\BibitemShut {NoStop}%
\bibitem [{\citenamefont {Ni}\ \emph {et~al.}(2010)\citenamefont {Ni},
  \citenamefont {Thaler}, \citenamefont {Yan}, \citenamefont {Kracher},
  \citenamefont {Colombier}, \citenamefont {Bud'ko}, \citenamefont {Canfield},\
  and\ \citenamefont {Hannahs}}]{Ni2010}%
  \BibitemOpen
  \bibfield  {author} {\bibinfo {author} {\bibfnamefont {N.}~\bibnamefont
  {Ni}}, \bibinfo {author} {\bibfnamefont {A.}~\bibnamefont {Thaler}}, \bibinfo
  {author} {\bibfnamefont {J.~Q.}\ \bibnamefont {Yan}}, \bibinfo {author}
  {\bibfnamefont {A.}~\bibnamefont {Kracher}}, \bibinfo {author} {\bibfnamefont
  {E.}~\bibnamefont {Colombier}}, \bibinfo {author} {\bibfnamefont {S.~L.}\
  \bibnamefont {Bud'ko}}, \bibinfo {author} {\bibfnamefont {P.~C.}\
  \bibnamefont {Canfield}}, \ and\ \bibinfo {author} {\bibfnamefont {S.~T.}\
  \bibnamefont {Hannahs}},\ }\href {\doibase 10.1103/PhysRevB.82.024519}
  {\bibfield  {journal} {\bibinfo  {journal} {Phys. Rev. B}\ }\textbf {\bibinfo
  {volume} {82}},\ \bibinfo {pages} {024519} (\bibinfo {year}
  {2010})}\BibitemShut {NoStop}%
\bibitem [{\citenamefont {Parker}\ \emph {et~al.}(2010)\citenamefont {Parker},
  \citenamefont {Smith}, \citenamefont {Lancaster}, \citenamefont {Steele},
  \citenamefont {Franke}, \citenamefont {Baker}, \citenamefont {Pratt},
  \citenamefont {Pitcher}, \citenamefont {Blundell},\ and\ \citenamefont
  {Clarke}}]{Parker2010}%
  \BibitemOpen
  \bibfield  {author} {\bibinfo {author} {\bibfnamefont {D.~R.}\ \bibnamefont
  {Parker}}, \bibinfo {author} {\bibfnamefont {M.~J.~P.}\ \bibnamefont
  {Smith}}, \bibinfo {author} {\bibfnamefont {T.}~\bibnamefont {Lancaster}},
  \bibinfo {author} {\bibfnamefont {A.~J.}\ \bibnamefont {Steele}}, \bibinfo
  {author} {\bibfnamefont {I.}~\bibnamefont {Franke}}, \bibinfo {author}
  {\bibfnamefont {P.~J.}\ \bibnamefont {Baker}}, \bibinfo {author}
  {\bibfnamefont {F.~L.}\ \bibnamefont {Pratt}}, \bibinfo {author}
  {\bibfnamefont {M.~J.}\ \bibnamefont {Pitcher}}, \bibinfo {author}
  {\bibfnamefont {S.~J.}\ \bibnamefont {Blundell}}, \ and\ \bibinfo {author}
  {\bibfnamefont {S.~J.}\ \bibnamefont {Clarke}},\ }\href {\doibase
  10.1103/PhysRevLett.104.057007} {\bibfield  {journal} {\bibinfo  {journal}
  {Phys. Rev. Lett.}\ }\textbf {\bibinfo {volume} {104}},\ \bibinfo {pages}
  {057007} (\bibinfo {year} {2010})}\BibitemShut {NoStop}%
\bibitem [{\citenamefont {Hassinger}\ \emph {et~al.}(2012)\citenamefont
  {Hassinger}, \citenamefont {Gredat}, \citenamefont {Valade}, \citenamefont
  {de~Cotret}, \citenamefont {Juneau-Fecteau}, \citenamefont {Reid},
  \citenamefont {Kim}, \citenamefont {Tanatar}, \citenamefont {Prozorov},
  \citenamefont {Shen}, \citenamefont {Wen}, \citenamefont {Doiron-Leyraud},\
  and\ \citenamefont {Taillefer}}]{Hassinger2012}%
  \BibitemOpen
  \bibfield  {author} {\bibinfo {author} {\bibfnamefont {E.}~\bibnamefont
  {Hassinger}}, \bibinfo {author} {\bibfnamefont {G.}~\bibnamefont {Gredat}},
  \bibinfo {author} {\bibfnamefont {F.}~\bibnamefont {Valade}}, \bibinfo
  {author} {\bibfnamefont {S.~R.}\ \bibnamefont {de~Cotret}}, \bibinfo {author}
  {\bibfnamefont {A.}~\bibnamefont {Juneau-Fecteau}}, \bibinfo {author}
  {\bibfnamefont {J.-P.}\ \bibnamefont {Reid}}, \bibinfo {author}
  {\bibfnamefont {H.}~\bibnamefont {Kim}}, \bibinfo {author} {\bibfnamefont
  {M.~A.}\ \bibnamefont {Tanatar}}, \bibinfo {author} {\bibfnamefont
  {R.}~\bibnamefont {Prozorov}}, \bibinfo {author} {\bibfnamefont
  {B.}~\bibnamefont {Shen}}, \bibinfo {author} {\bibfnamefont {H.-H.}\
  \bibnamefont {Wen}}, \bibinfo {author} {\bibfnamefont {N.}~\bibnamefont
  {Doiron-Leyraud}}, \ and\ \bibinfo {author} {\bibfnamefont {L.}~\bibnamefont
  {Taillefer}},\ }\href {\doibase 10.1103/PhysRevB.86.140502} {\bibfield
  {journal} {\bibinfo  {journal} {Phys. Rev. B}\ }\textbf {\bibinfo {volume}
  {86}},\ \bibinfo {pages} {140502} (\bibinfo {year} {2012})}\BibitemShut
  {NoStop}%
\bibitem [{\citenamefont {Avci}\ \emph {et~al.}(2014)\citenamefont {Avci},
  \citenamefont {Chmaissem}, \citenamefont {Allred}, \citenamefont
  {Rosenkranz}, \citenamefont {Eremin}, \citenamefont {Chubukov}, \citenamefont
  {Bugaris}, \citenamefont {Chung}, \citenamefont {Kanatzidis}, \citenamefont
  {Castellan}, \citenamefont {Schlueter}, \citenamefont {Claus}, \citenamefont
  {Khalyavin}, \citenamefont {Manuel}, \citenamefont {Daoud-Aladine},\ and\
  \citenamefont {Osborn}}]{Avci2014}%
  \BibitemOpen
  \bibfield  {author} {\bibinfo {author} {\bibfnamefont {S.}~\bibnamefont
  {Avci}}, \bibinfo {author} {\bibfnamefont {O.}~\bibnamefont {Chmaissem}},
  \bibinfo {author} {\bibfnamefont {J.}~\bibnamefont {Allred}}, \bibinfo
  {author} {\bibfnamefont {S.}~\bibnamefont {Rosenkranz}}, \bibinfo {author}
  {\bibfnamefont {I.}~\bibnamefont {Eremin}}, \bibinfo {author} {\bibfnamefont
  {A.}~\bibnamefont {Chubukov}}, \bibinfo {author} {\bibfnamefont
  {D.}~\bibnamefont {Bugaris}}, \bibinfo {author} {\bibfnamefont
  {D.}~\bibnamefont {Chung}}, \bibinfo {author} {\bibfnamefont
  {M.}~\bibnamefont {Kanatzidis}}, \bibinfo {author} {\bibfnamefont {J.-P.}\
  \bibnamefont {Castellan}}, \bibinfo {author} {\bibfnamefont {J.}~\bibnamefont
  {Schlueter}}, \bibinfo {author} {\bibfnamefont {H.}~\bibnamefont {Claus}},
  \bibinfo {author} {\bibfnamefont {D.}~\bibnamefont {Khalyavin}}, \bibinfo
  {author} {\bibfnamefont {P.}~\bibnamefont {Manuel}}, \bibinfo {author}
  {\bibfnamefont {A.}~\bibnamefont {Daoud-Aladine}}, \ and\ \bibinfo {author}
  {\bibfnamefont {R.}~\bibnamefont {Osborn}},\ }\href {\doibase
  10.1038/ncomms4845} {\bibfield  {journal} {\bibinfo  {journal} {Nature
  Communications}\ }\textbf {\bibinfo {volume} {5}},\ \bibinfo {pages} {3845}
  (\bibinfo {year} {2014})}\BibitemShut {NoStop}%
\bibitem [{\citenamefont {B{\"o}hmer}\ \emph {et~al.}(2015)\citenamefont
  {B{\"o}hmer}, \citenamefont {Hardy}, \citenamefont {Wang}, \citenamefont
  {Wolf}, \citenamefont {Schweiss},\ and\ \citenamefont
  {Meingast}}]{Boehmer2015II}%
  \BibitemOpen
  \bibfield  {author} {\bibinfo {author} {\bibfnamefont {A.~E.}\ \bibnamefont
  {B{\"o}hmer}}, \bibinfo {author} {\bibfnamefont {F.}~\bibnamefont {Hardy}},
  \bibinfo {author} {\bibfnamefont {L.}~\bibnamefont {Wang}}, \bibinfo {author}
  {\bibfnamefont {T.}~\bibnamefont {Wolf}}, \bibinfo {author} {\bibfnamefont
  {P.}~\bibnamefont {Schweiss}}, \ and\ \bibinfo {author} {\bibfnamefont
  {C.}~\bibnamefont {Meingast}},\ }\href {\doibase 10.1038/ncomms8911}
  {\bibfield  {journal} {\bibinfo  {journal} {Nature Communications}\ }\textbf
  {\bibinfo {volume} {6}},\ \bibinfo {pages} {8911} (\bibinfo {year}
  {2015})}\BibitemShut {NoStop}%
\bibitem [{\citenamefont {Allred}\ \emph {et~al.}(2016)\citenamefont {Allred},
  \citenamefont {Taddei}, \citenamefont {Bugaris}, \citenamefont {Krogstad},
  \citenamefont {Lapidus}, \citenamefont {Chung}, \citenamefont {Claus},
  \citenamefont {Kanatzidis}, \citenamefont {Brown}, \citenamefont {Kang},
  \citenamefont {Fernandes}, \citenamefont {Eremin}, \citenamefont
  {Rosenkranz}, \citenamefont {Chmaissem},\ and\ \citenamefont
  {Osborn}}]{Allred2016}%
  \BibitemOpen
  \bibfield  {author} {\bibinfo {author} {\bibfnamefont {J.~M.}\ \bibnamefont
  {Allred}}, \bibinfo {author} {\bibfnamefont {K.~M.}\ \bibnamefont {Taddei}},
  \bibinfo {author} {\bibfnamefont {D.~E.}\ \bibnamefont {Bugaris}}, \bibinfo
  {author} {\bibfnamefont {M.~J.}\ \bibnamefont {Krogstad}}, \bibinfo {author}
  {\bibfnamefont {S.~H.}\ \bibnamefont {Lapidus}}, \bibinfo {author}
  {\bibfnamefont {D.~Y.}\ \bibnamefont {Chung}}, \bibinfo {author}
  {\bibfnamefont {H.}~\bibnamefont {Claus}}, \bibinfo {author} {\bibfnamefont
  {M.}~\bibnamefont {Kanatzidis}}, \bibinfo {author} {\bibfnamefont {D.~E.}\
  \bibnamefont {Brown}}, \bibinfo {author} {\bibfnamefont {J.}~\bibnamefont
  {Kang}}, \bibinfo {author} {\bibfnamefont {R.~M.}\ \bibnamefont {Fernandes}},
  \bibinfo {author} {\bibfnamefont {I.}~\bibnamefont {Eremin}}, \bibinfo
  {author} {\bibfnamefont {S.}~\bibnamefont {Rosenkranz}}, \bibinfo {author}
  {\bibfnamefont {O.}~\bibnamefont {Chmaissem}}, \ and\ \bibinfo {author}
  {\bibfnamefont {R.}~\bibnamefont {Osborn}},\ }\href
  {http://dx.doi.org/10.1038/nphys3629} {\bibfield  {journal} {\bibinfo
  {journal} {Nature Physics}\ }\textbf {\bibinfo {volume} {12}} (\bibinfo
  {year} {2016})}\BibitemShut {NoStop}%
\bibitem [{\citenamefont {Kreyssig}\ \emph {et~al.}(2010)\citenamefont
  {Kreyssig}, \citenamefont {Kim}, \citenamefont {Nandi}, \citenamefont
  {Pratt}, \citenamefont {Tian}, \citenamefont {Zarestky}, \citenamefont {Ni},
  \citenamefont {Thaler}, \citenamefont {Bud'ko}, \citenamefont {Canfield},
  \citenamefont {McQueeney},\ and\ \citenamefont {Goldman}}]{Kreyssig2010}%
  \BibitemOpen
  \bibfield  {author} {\bibinfo {author} {\bibfnamefont {A.}~\bibnamefont
  {Kreyssig}}, \bibinfo {author} {\bibfnamefont {M.~G.}\ \bibnamefont {Kim}},
  \bibinfo {author} {\bibfnamefont {S.}~\bibnamefont {Nandi}}, \bibinfo
  {author} {\bibfnamefont {D.~K.}\ \bibnamefont {Pratt}}, \bibinfo {author}
  {\bibfnamefont {W.}~\bibnamefont {Tian}}, \bibinfo {author} {\bibfnamefont
  {J.~L.}\ \bibnamefont {Zarestky}}, \bibinfo {author} {\bibfnamefont
  {N.}~\bibnamefont {Ni}}, \bibinfo {author} {\bibfnamefont {A.}~\bibnamefont
  {Thaler}}, \bibinfo {author} {\bibfnamefont {S.~L.}\ \bibnamefont {Bud'ko}},
  \bibinfo {author} {\bibfnamefont {P.~C.}\ \bibnamefont {Canfield}}, \bibinfo
  {author} {\bibfnamefont {R.~J.}\ \bibnamefont {McQueeney}}, \ and\ \bibinfo
  {author} {\bibfnamefont {A.~I.}\ \bibnamefont {Goldman}},\ }\href {\doibase
  10.1103/PhysRevB.81.134512} {\bibfield  {journal} {\bibinfo  {journal} {Phys.
  Rev. B}\ }\textbf {\bibinfo {volume} {81}},\ \bibinfo {pages} {134512}
  (\bibinfo {year} {2010})}\BibitemShut {NoStop}%
\bibitem [{\citenamefont {Allred}\ \emph {et~al.}(2015)\citenamefont {Allred},
  \citenamefont {Avci}, \citenamefont {Chung}, \citenamefont {Claus},
  \citenamefont {Khalyavin}, \citenamefont {Manuel}, \citenamefont {Taddei},
  \citenamefont {Kanatzidis}, \citenamefont {Rosenkranz}, \citenamefont
  {Osborn},\ and\ \citenamefont {Chmaissem}}]{Taddei2015}%
  \BibitemOpen
  \bibfield  {author} {\bibinfo {author} {\bibfnamefont {J.~M.}\ \bibnamefont
  {Allred}}, \bibinfo {author} {\bibfnamefont {S.}~\bibnamefont {Avci}},
  \bibinfo {author} {\bibfnamefont {D.~Y.}\ \bibnamefont {Chung}}, \bibinfo
  {author} {\bibfnamefont {H.}~\bibnamefont {Claus}}, \bibinfo {author}
  {\bibfnamefont {D.~D.}\ \bibnamefont {Khalyavin}}, \bibinfo {author}
  {\bibfnamefont {P.}~\bibnamefont {Manuel}}, \bibinfo {author} {\bibfnamefont
  {K.~M.}\ \bibnamefont {Taddei}}, \bibinfo {author} {\bibfnamefont {M.~G.}\
  \bibnamefont {Kanatzidis}}, \bibinfo {author} {\bibfnamefont
  {S.}~\bibnamefont {Rosenkranz}}, \bibinfo {author} {\bibfnamefont
  {R.}~\bibnamefont {Osborn}}, \ and\ \bibinfo {author} {\bibfnamefont
  {O.}~\bibnamefont {Chmaissem}},\ }\href {\doibase 10.1103/PhysRevB.92.094515}
  {\bibfield  {journal} {\bibinfo  {journal} {Phys. Rev. B}\ }\textbf {\bibinfo
  {volume} {92}},\ \bibinfo {pages} {094515} (\bibinfo {year}
  {2015})}\BibitemShut {NoStop}%
\bibitem [{\citenamefont {Hassinger}\ \emph {et~al.}(2016)\citenamefont
  {Hassinger}, \citenamefont {Gredat}, \citenamefont {Valade}, \citenamefont
  {de~Cotret}, \citenamefont {Cyr-Choini\`ere}, \citenamefont {Juneau-Fecteau},
  \citenamefont {Reid}, \citenamefont {Kim}, \citenamefont {Tanatar},
  \citenamefont {Prozorov}, \citenamefont {Shen}, \citenamefont {Wen},
  \citenamefont {Doiron-Leyraud},\ and\ \citenamefont
  {Taillefer}}]{Hassinger_2016}%
  \BibitemOpen
  \bibfield  {author} {\bibinfo {author} {\bibfnamefont {E.}~\bibnamefont
  {Hassinger}}, \bibinfo {author} {\bibfnamefont {G.}~\bibnamefont {Gredat}},
  \bibinfo {author} {\bibfnamefont {F.}~\bibnamefont {Valade}}, \bibinfo
  {author} {\bibfnamefont {S.~R.}\ \bibnamefont {de~Cotret}}, \bibinfo {author}
  {\bibfnamefont {O.}~\bibnamefont {Cyr-Choini\`ere}}, \bibinfo {author}
  {\bibfnamefont {A.}~\bibnamefont {Juneau-Fecteau}}, \bibinfo {author}
  {\bibfnamefont {J.-P.}\ \bibnamefont {Reid}}, \bibinfo {author}
  {\bibfnamefont {H.}~\bibnamefont {Kim}}, \bibinfo {author} {\bibfnamefont
  {M.~A.}\ \bibnamefont {Tanatar}}, \bibinfo {author} {\bibfnamefont
  {R.}~\bibnamefont {Prozorov}}, \bibinfo {author} {\bibfnamefont
  {B.}~\bibnamefont {Shen}}, \bibinfo {author} {\bibfnamefont {H.-H.}\
  \bibnamefont {Wen}}, \bibinfo {author} {\bibfnamefont {N.}~\bibnamefont
  {Doiron-Leyraud}}, \ and\ \bibinfo {author} {\bibfnamefont {L.}~\bibnamefont
  {Taillefer}},\ }\href {\doibase 10.1103/PhysRevB.93.144401} {\bibfield
  {journal} {\bibinfo  {journal} {Phys. Rev. B}\ }\textbf {\bibinfo {volume}
  {93}},\ \bibinfo {pages} {144401} (\bibinfo {year} {2016})}\BibitemShut
  {NoStop}%
\bibitem [{\citenamefont {Wang}\ \emph {et~al.}(2017)\citenamefont {Wang},
  \citenamefont {Hardy}, \citenamefont {Wolf}, \citenamefont {Adelmann},
  \citenamefont {Fromknecht}, \citenamefont {Schweiss},\ and\ \citenamefont
  {Meingast}}]{Wang2017II}%
  \BibitemOpen
  \bibfield  {author} {\bibinfo {author} {\bibfnamefont {L.}~\bibnamefont
  {Wang}}, \bibinfo {author} {\bibfnamefont {F.}~\bibnamefont {Hardy}},
  \bibinfo {author} {\bibfnamefont {T.}~\bibnamefont {Wolf}}, \bibinfo {author}
  {\bibfnamefont {P.}~\bibnamefont {Adelmann}}, \bibinfo {author}
  {\bibfnamefont {R.}~\bibnamefont {Fromknecht}}, \bibinfo {author}
  {\bibfnamefont {P.}~\bibnamefont {Schweiss}}, \ and\ \bibinfo {author}
  {\bibfnamefont {C.}~\bibnamefont {Meingast}},\ }\href {\doibase
  10.1002/pssb.201600153} {\bibfield  {journal} {\bibinfo  {journal} {physica
  status solidi (b)}\ }\textbf {\bibinfo {volume} {254}},\ \bibinfo {pages}
  {1600153} (\bibinfo {year} {2017})}\BibitemShut {NoStop}%
\bibitem [{\citenamefont {Taddei}\ \emph {et~al.}(2016)\citenamefont {Taddei},
  \citenamefont {Allred}, \citenamefont {Bugaris}, \citenamefont {Lapidus},
  \citenamefont {Krogstad}, \citenamefont {Stadel}, \citenamefont {Claus},
  \citenamefont {Chung}, \citenamefont {Kanatzidis}, \citenamefont
  {Rosenkranz}, \citenamefont {Osborn},\ and\ \citenamefont
  {Chmaissem}}]{Taddei2016}%
  \BibitemOpen
  \bibfield  {author} {\bibinfo {author} {\bibfnamefont {K.~M.}\ \bibnamefont
  {Taddei}}, \bibinfo {author} {\bibfnamefont {J.~M.}\ \bibnamefont {Allred}},
  \bibinfo {author} {\bibfnamefont {D.~E.}\ \bibnamefont {Bugaris}}, \bibinfo
  {author} {\bibfnamefont {S.}~\bibnamefont {Lapidus}}, \bibinfo {author}
  {\bibfnamefont {M.~J.}\ \bibnamefont {Krogstad}}, \bibinfo {author}
  {\bibfnamefont {R.}~\bibnamefont {Stadel}}, \bibinfo {author} {\bibfnamefont
  {H.}~\bibnamefont {Claus}}, \bibinfo {author} {\bibfnamefont {D.~Y.}\
  \bibnamefont {Chung}}, \bibinfo {author} {\bibfnamefont {M.~G.}\ \bibnamefont
  {Kanatzidis}}, \bibinfo {author} {\bibfnamefont {S.}~\bibnamefont
  {Rosenkranz}}, \bibinfo {author} {\bibfnamefont {R.}~\bibnamefont {Osborn}},
  \ and\ \bibinfo {author} {\bibfnamefont {O.}~\bibnamefont {Chmaissem}},\
  }\href {\doibase 10.1103/PhysRevB.93.134510} {\bibfield  {journal} {\bibinfo
  {journal} {Phys. Rev. B}\ }\textbf {\bibinfo {volume} {93}},\ \bibinfo
  {pages} {134510} (\bibinfo {year} {2016})}\BibitemShut {NoStop}%
\bibitem [{\citenamefont {Taddei}\ \emph {et~al.}(2017)\citenamefont {Taddei},
  \citenamefont {Allred}, \citenamefont {Bugaris}, \citenamefont {Lapidus},
  \citenamefont {Krogstad}, \citenamefont {Claus}, \citenamefont {Chung},
  \citenamefont {Kanatzidis}, \citenamefont {Osborn}, \citenamefont
  {Rosenkranz},\ and\ \citenamefont {Chmaissem}}]{Taddei2017}%
  \BibitemOpen
  \bibfield  {author} {\bibinfo {author} {\bibfnamefont {K.~M.}\ \bibnamefont
  {Taddei}}, \bibinfo {author} {\bibfnamefont {J.~M.}\ \bibnamefont {Allred}},
  \bibinfo {author} {\bibfnamefont {D.~E.}\ \bibnamefont {Bugaris}}, \bibinfo
  {author} {\bibfnamefont {S.~H.}\ \bibnamefont {Lapidus}}, \bibinfo {author}
  {\bibfnamefont {M.~J.}\ \bibnamefont {Krogstad}}, \bibinfo {author}
  {\bibfnamefont {H.}~\bibnamefont {Claus}}, \bibinfo {author} {\bibfnamefont
  {D.~Y.}\ \bibnamefont {Chung}}, \bibinfo {author} {\bibfnamefont {M.~G.}\
  \bibnamefont {Kanatzidis}}, \bibinfo {author} {\bibfnamefont
  {R.}~\bibnamefont {Osborn}}, \bibinfo {author} {\bibfnamefont
  {S.}~\bibnamefont {Rosenkranz}}, \ and\ \bibinfo {author} {\bibfnamefont
  {O.}~\bibnamefont {Chmaissem}},\ }\href {\doibase 10.1103/PhysRevB.95.064508}
  {\bibfield  {journal} {\bibinfo  {journal} {Phys. Rev. B}\ }\textbf {\bibinfo
  {volume} {95}},\ \bibinfo {pages} {064508} (\bibinfo {year}
  {2017})}\BibitemShut {NoStop}%
\bibitem [{\citenamefont {Meier}\ \emph {et~al.}(2018)\citenamefont {Meier},
  \citenamefont {Ding}, \citenamefont {Kreyssig}, \citenamefont {Bud’ko},
  \citenamefont {Sapkota}, \citenamefont {Kothapalli}, \citenamefont {Borisov},
  \citenamefont {Valent{\'i}}, \citenamefont {Batista}, \citenamefont {Orth},
  \citenamefont {Fernandes}, \citenamefont {Goldman}, \citenamefont {Furukawa},
  \citenamefont {B{\"o}hmer},\ and\ \citenamefont {Canfield}}]{Meier2018}%
  \BibitemOpen
  \bibfield  {author} {\bibinfo {author} {\bibfnamefont {W.~R.}\ \bibnamefont
  {Meier}}, \bibinfo {author} {\bibfnamefont {Q.-P.}\ \bibnamefont {Ding}},
  \bibinfo {author} {\bibfnamefont {A.}~\bibnamefont {Kreyssig}}, \bibinfo
  {author} {\bibfnamefont {S.~L.}\ \bibnamefont {Bud’ko}}, \bibinfo {author}
  {\bibfnamefont {A.}~\bibnamefont {Sapkota}}, \bibinfo {author} {\bibfnamefont
  {K.}~\bibnamefont {Kothapalli}}, \bibinfo {author} {\bibfnamefont
  {V.}~\bibnamefont {Borisov}}, \bibinfo {author} {\bibfnamefont
  {R.}~\bibnamefont {Valent{\'i}}}, \bibinfo {author} {\bibfnamefont {C.~D.}\
  \bibnamefont {Batista}}, \bibinfo {author} {\bibfnamefont {P.~P.}\
  \bibnamefont {Orth}}, \bibinfo {author} {\bibfnamefont {R.~M.}\ \bibnamefont
  {Fernandes}}, \bibinfo {author} {\bibfnamefont {A.~I.}\ \bibnamefont
  {Goldman}}, \bibinfo {author} {\bibfnamefont {Y.}~\bibnamefont {Furukawa}},
  \bibinfo {author} {\bibfnamefont {A.~E.}\ \bibnamefont {B{\"o}hmer}}, \ and\
  \bibinfo {author} {\bibfnamefont {P.~C.}\ \bibnamefont {Canfield}},\ }\href
  {\doibase 10.1038/s41535-017-0076-x} {\bibfield  {journal} {\bibinfo
  {journal} {npj Quantum Materials}\ }\textbf {\bibinfo {volume} {3}},\
  \bibinfo {pages} {5} (\bibinfo {year} {2018})}\BibitemShut {NoStop}%
\bibitem [{\citenamefont {McQueen}\ \emph {et~al.}(2009)\citenamefont
  {McQueen}, \citenamefont {Williams}, \citenamefont {Stephens}, \citenamefont
  {Tao}, \citenamefont {Zhu}, \citenamefont {Ksenofontov}, \citenamefont
  {Casper}, \citenamefont {Felser},\ and\ \citenamefont {Cava}}]{McQueen2009}%
  \BibitemOpen
  \bibfield  {author} {\bibinfo {author} {\bibfnamefont {T.~M.}\ \bibnamefont
  {McQueen}}, \bibinfo {author} {\bibfnamefont {A.~J.}\ \bibnamefont
  {Williams}}, \bibinfo {author} {\bibfnamefont {P.~W.}\ \bibnamefont
  {Stephens}}, \bibinfo {author} {\bibfnamefont {J.}~\bibnamefont {Tao}},
  \bibinfo {author} {\bibfnamefont {Y.}~\bibnamefont {Zhu}}, \bibinfo {author}
  {\bibfnamefont {V.}~\bibnamefont {Ksenofontov}}, \bibinfo {author}
  {\bibfnamefont {F.}~\bibnamefont {Casper}}, \bibinfo {author} {\bibfnamefont
  {C.}~\bibnamefont {Felser}}, \ and\ \bibinfo {author} {\bibfnamefont {R.~J.}\
  \bibnamefont {Cava}},\ }\href {\doibase 10.1103/PhysRevLett.103.057002}
  {\bibfield  {journal} {\bibinfo  {journal} {Phys. Rev. Lett.}\ }\textbf
  {\bibinfo {volume} {103}},\ \bibinfo {pages} {057002} (\bibinfo {year}
  {2009})}\BibitemShut {NoStop}%
\bibitem [{\citenamefont {Watson}\ \emph {et~al.}(2015)\citenamefont {Watson},
  \citenamefont {Kim}, \citenamefont {Haghighirad}, \citenamefont {Davies},
  \citenamefont {McCollam}, \citenamefont {Narayanan}, \citenamefont {Blake},
  \citenamefont {Chen}, \citenamefont {Ghannadzadeh}, \citenamefont
  {Schofield}, \citenamefont {Hoesch}, \citenamefont {Meingast}, \citenamefont
  {Wolf},\ and\ \citenamefont {Coldea}}]{Watson2015}%
  \BibitemOpen
  \bibfield  {author} {\bibinfo {author} {\bibfnamefont {M.~D.}\ \bibnamefont
  {Watson}}, \bibinfo {author} {\bibfnamefont {T.~K.}\ \bibnamefont {Kim}},
  \bibinfo {author} {\bibfnamefont {A.~A.}\ \bibnamefont {Haghighirad}},
  \bibinfo {author} {\bibfnamefont {N.~R.}\ \bibnamefont {Davies}}, \bibinfo
  {author} {\bibfnamefont {A.}~\bibnamefont {McCollam}}, \bibinfo {author}
  {\bibfnamefont {A.}~\bibnamefont {Narayanan}}, \bibinfo {author}
  {\bibfnamefont {S.~F.}\ \bibnamefont {Blake}}, \bibinfo {author}
  {\bibfnamefont {Y.~L.}\ \bibnamefont {Chen}}, \bibinfo {author}
  {\bibfnamefont {S.}~\bibnamefont {Ghannadzadeh}}, \bibinfo {author}
  {\bibfnamefont {A.~J.}\ \bibnamefont {Schofield}}, \bibinfo {author}
  {\bibfnamefont {M.}~\bibnamefont {Hoesch}}, \bibinfo {author} {\bibfnamefont
  {C.}~\bibnamefont {Meingast}}, \bibinfo {author} {\bibfnamefont
  {T.}~\bibnamefont {Wolf}}, \ and\ \bibinfo {author} {\bibfnamefont {A.~I.}\
  \bibnamefont {Coldea}},\ }\href {\doibase 10.1103/PhysRevB.91.155106}
  {\bibfield  {journal} {\bibinfo  {journal} {Phys. Rev. B}\ }\textbf {\bibinfo
  {volume} {91}},\ \bibinfo {pages} {155106} (\bibinfo {year}
  {2015})}\BibitemShut {NoStop}%
\bibitem [{\citenamefont {{Wang}}\ \emph {et~al.}(2016)\citenamefont {{Wang}},
  \citenamefont {{Shen}}, \citenamefont {{Pan}}, \citenamefont {{Hao}},
  \citenamefont {{Ma}}, \citenamefont {{Zhou}}, \citenamefont {{Steffens}},
  \citenamefont {{Schmalzl}}, \citenamefont {{Forrest}}, \citenamefont
  {{Abdel-Hafiez}}, \citenamefont {Chen}, \citenamefont {{Chareev}},
  \citenamefont {{Vasiliev}}, \citenamefont {{Bourges}}, \citenamefont
  {{Sidis}}, \citenamefont {{Cao}},\ and\ \citenamefont {{Zhao}}}]{Wang2015}%
  \BibitemOpen
  \bibfield  {author} {\bibinfo {author} {\bibfnamefont {Q.}~\bibnamefont
  {{Wang}}}, \bibinfo {author} {\bibfnamefont {Y.}~\bibnamefont {{Shen}}},
  \bibinfo {author} {\bibfnamefont {B.}~\bibnamefont {{Pan}}}, \bibinfo
  {author} {\bibfnamefont {Y.}~\bibnamefont {{Hao}}}, \bibinfo {author}
  {\bibfnamefont {M.}~\bibnamefont {{Ma}}}, \bibinfo {author} {\bibfnamefont
  {F.}~\bibnamefont {{Zhou}}}, \bibinfo {author} {\bibfnamefont
  {P.}~\bibnamefont {{Steffens}}}, \bibinfo {author} {\bibfnamefont
  {K.}~\bibnamefont {{Schmalzl}}}, \bibinfo {author} {\bibfnamefont {T.~R.}\
  \bibnamefont {{Forrest}}}, \bibinfo {author} {\bibfnamefont {M.}~\bibnamefont
  {{Abdel-Hafiez}}}, \bibinfo {author} {\bibfnamefont {X.}~\bibnamefont
  {Chen}}, \bibinfo {author} {\bibfnamefont {D.~A.}\ \bibnamefont {{Chareev}}},
  \bibinfo {author} {\bibfnamefont {A.~N.}\ \bibnamefont {{Vasiliev}}},
  \bibinfo {author} {\bibfnamefont {P.}~\bibnamefont {{Bourges}}}, \bibinfo
  {author} {\bibfnamefont {Y.}~\bibnamefont {{Sidis}}}, \bibinfo {author}
  {\bibfnamefont {H.}~\bibnamefont {{Cao}}}, \ and\ \bibinfo {author}
  {\bibfnamefont {J.}~\bibnamefont {{Zhao}}},\ }\href {\doibase
  10.1038/nmat4492} {\bibfield  {journal} {\bibinfo  {journal} {Nature
  Materials}\ }\textbf {\bibinfo {volume} {15}},\ \bibinfo {pages} {159}
  (\bibinfo {year} {2016})}\BibitemShut {NoStop}%
\bibitem [{\citenamefont {Fanfarillo}\ \emph {et~al.}(2016)\citenamefont
  {Fanfarillo}, \citenamefont {Mansart}, \citenamefont {Toulemonde},
  \citenamefont {Cercellier}, \citenamefont {Le~F\`evre}, \citenamefont
  {Bertran}, \citenamefont {Valenzuela}, \citenamefont {Benfatto},\ and\
  \citenamefont {Brouet}}]{Fanfarillo2016}%
  \BibitemOpen
  \bibfield  {author} {\bibinfo {author} {\bibfnamefont {L.}~\bibnamefont
  {Fanfarillo}}, \bibinfo {author} {\bibfnamefont {J.}~\bibnamefont {Mansart}},
  \bibinfo {author} {\bibfnamefont {P.}~\bibnamefont {Toulemonde}}, \bibinfo
  {author} {\bibfnamefont {H.}~\bibnamefont {Cercellier}}, \bibinfo {author}
  {\bibfnamefont {P.}~\bibnamefont {Le~F\`evre}}, \bibinfo {author}
  {\bibfnamefont {F.}~\bibnamefont {Bertran}}, \bibinfo {author} {\bibfnamefont
  {B.}~\bibnamefont {Valenzuela}}, \bibinfo {author} {\bibfnamefont
  {L.}~\bibnamefont {Benfatto}}, \ and\ \bibinfo {author} {\bibfnamefont
  {V.}~\bibnamefont {Brouet}},\ }\href {\doibase 10.1103/PhysRevB.94.155138}
  {\bibfield  {journal} {\bibinfo  {journal} {Phys. Rev. B}\ }\textbf {\bibinfo
  {volume} {94}},\ \bibinfo {pages} {155138} (\bibinfo {year}
  {2016})}\BibitemShut {NoStop}%
\bibitem [{\citenamefont {Yamakawa}\ \emph {et~al.}(2016)\citenamefont
  {Yamakawa}, \citenamefont {Onari},\ and\ \citenamefont
  {Kontani}}]{Yamakawa_PRX_16}%
  \BibitemOpen
  \bibfield  {author} {\bibinfo {author} {\bibfnamefont {Y.}~\bibnamefont
  {Yamakawa}}, \bibinfo {author} {\bibfnamefont {S.}~\bibnamefont {Onari}}, \
  and\ \bibinfo {author} {\bibfnamefont {H.}~\bibnamefont {Kontani}},\ }\href
  {\doibase 10.1103/PhysRevX.6.021032} {\bibfield  {journal} {\bibinfo
  {journal} {Phys. Rev. X}\ }\textbf {\bibinfo {volume} {6}},\ \bibinfo {pages}
  {021032} (\bibinfo {year} {2016})}\BibitemShut {NoStop}%
\bibitem [{\citenamefont {Tanatar}\ \emph {et~al.}(2016)\citenamefont
  {Tanatar}, \citenamefont {B\"ohmer}, \citenamefont {Timmons}, \citenamefont
  {Sch\"utt}, \citenamefont {Drachuck}, \citenamefont {Taufour}, \citenamefont
  {Kothapalli}, \citenamefont {Kreyssig}, \citenamefont {Bud'ko}, \citenamefont
  {Canfield}, \citenamefont {Fernandes},\ and\ \citenamefont
  {Prozorov}}]{Tanatar2016}%
  \BibitemOpen
  \bibfield  {author} {\bibinfo {author} {\bibfnamefont {M.~A.}\ \bibnamefont
  {Tanatar}}, \bibinfo {author} {\bibfnamefont {A.~E.}\ \bibnamefont
  {B\"ohmer}}, \bibinfo {author} {\bibfnamefont {E.~I.}\ \bibnamefont
  {Timmons}}, \bibinfo {author} {\bibfnamefont {M.}~\bibnamefont {Sch\"utt}},
  \bibinfo {author} {\bibfnamefont {G.}~\bibnamefont {Drachuck}}, \bibinfo
  {author} {\bibfnamefont {V.}~\bibnamefont {Taufour}}, \bibinfo {author}
  {\bibfnamefont {K.}~\bibnamefont {Kothapalli}}, \bibinfo {author}
  {\bibfnamefont {A.}~\bibnamefont {Kreyssig}}, \bibinfo {author}
  {\bibfnamefont {S.~L.}\ \bibnamefont {Bud'ko}}, \bibinfo {author}
  {\bibfnamefont {P.~C.}\ \bibnamefont {Canfield}}, \bibinfo {author}
  {\bibfnamefont {R.~M.}\ \bibnamefont {Fernandes}}, \ and\ \bibinfo {author}
  {\bibfnamefont {R.}~\bibnamefont {Prozorov}},\ }\href {\doibase
  10.1103/PhysRevLett.117.127001} {\bibfield  {journal} {\bibinfo  {journal}
  {Phys. Rev. Lett.}\ }\textbf {\bibinfo {volume} {117}},\ \bibinfo {pages}
  {127001} (\bibinfo {year} {2016})}\BibitemShut {NoStop}%
\bibitem [{\citenamefont {Chinotti}\ \emph {et~al.}(2017)\citenamefont
  {Chinotti}, \citenamefont {Pal}, \citenamefont {Degiorgi}, \citenamefont
  {B\"ohmer},\ and\ \citenamefont {Canfield}}]{Chinotti2017}%
  \BibitemOpen
  \bibfield  {author} {\bibinfo {author} {\bibfnamefont {M.}~\bibnamefont
  {Chinotti}}, \bibinfo {author} {\bibfnamefont {A.}~\bibnamefont {Pal}},
  \bibinfo {author} {\bibfnamefont {L.}~\bibnamefont {Degiorgi}}, \bibinfo
  {author} {\bibfnamefont {A.~E.}\ \bibnamefont {B\"ohmer}}, \ and\ \bibinfo
  {author} {\bibfnamefont {P.~C.}\ \bibnamefont {Canfield}},\ }\href {\doibase
  10.1103/PhysRevB.96.121112} {\bibfield  {journal} {\bibinfo  {journal} {Phys.
  Rev. B}\ }\textbf {\bibinfo {volume} {96}},\ \bibinfo {pages} {121112}
  (\bibinfo {year} {2017})}\BibitemShut {NoStop}%
\bibitem [{\citenamefont {Watson}\ \emph {et~al.}(2017)\citenamefont {Watson},
  \citenamefont {Haghighirad}, \citenamefont {Rhodes}, \citenamefont {Hoesch},\
  and\ \citenamefont {Kim}}]{Watson2017II}%
  \BibitemOpen
  \bibfield  {author} {\bibinfo {author} {\bibfnamefont {M.~D.}\ \bibnamefont
  {Watson}}, \bibinfo {author} {\bibfnamefont {A.~A.}\ \bibnamefont
  {Haghighirad}}, \bibinfo {author} {\bibfnamefont {L.~C.}\ \bibnamefont
  {Rhodes}}, \bibinfo {author} {\bibfnamefont {M.}~\bibnamefont {Hoesch}}, \
  and\ \bibinfo {author} {\bibfnamefont {T.~K.}\ \bibnamefont {Kim}},\ }\href
  {http://stacks.iop.org/1367-2630/19/i=10/a=103021} {\bibfield  {journal}
  {\bibinfo  {journal} {New Journal of Physics}\ }\textbf {\bibinfo {volume}
  {19}},\ \bibinfo {pages} {103021} (\bibinfo {year} {2017})}\BibitemShut
  {NoStop}%
\bibitem [{\citenamefont {{He}}\ \emph {et~al.}(2017)\citenamefont {{He}},
  \citenamefont {{Wang}}, \citenamefont {{Hardy}}, \citenamefont {{Xu}},
  \citenamefont {{Wolf}}, \citenamefont {{Adelmann}},\ and\ \citenamefont
  {{Meingast}}}]{He2017}%
  \BibitemOpen
  \bibfield  {author} {\bibinfo {author} {\bibfnamefont {M.}~\bibnamefont
  {{He}}}, \bibinfo {author} {\bibfnamefont {L.}~\bibnamefont {{Wang}}},
  \bibinfo {author} {\bibfnamefont {F.}~\bibnamefont {{Hardy}}}, \bibinfo
  {author} {\bibfnamefont {L.}~\bibnamefont {{Xu}}}, \bibinfo {author}
  {\bibfnamefont {T.}~\bibnamefont {{Wolf}}}, \bibinfo {author} {\bibfnamefont
  {P.}~\bibnamefont {{Adelmann}}}, \ and\ \bibinfo {author} {\bibfnamefont
  {C.}~\bibnamefont {{Meingast}}},\ }\href
  {http://adsabs.harvard.edu/abs/2017arXiv170903861H} {\bibfield  {journal}
  {\bibinfo  {journal} {ArXiv e-prints}\ } (\bibinfo {year} {2017})},\ \Eprint
  {http://arxiv.org/abs/1709.03861} {1709.03861} \BibitemShut {NoStop}%
\bibitem [{\citenamefont {B\"{o}hmer}\ and\ \citenamefont
  {Kreisel}(2018)}]{Boehmer2017}%
  \BibitemOpen
  \bibfield  {author} {\bibinfo {author} {\bibfnamefont {A.~E.}\ \bibnamefont
  {B\"{o}hmer}}\ and\ \bibinfo {author} {\bibfnamefont {A.}~\bibnamefont
  {Kreisel}},\ }\href {\doibase 10.1088/1361-648X/aa9caa} {\bibfield  {journal}
  {\bibinfo  {journal} {Journal of Physics: Condensed Matter}\ }\textbf
  {\bibinfo {volume} {30}},\ \bibinfo {pages} {023001} (\bibinfo {year}
  {2018})}\BibitemShut {NoStop}%
\bibitem [{\citenamefont {Bendele}\ \emph {et~al.}(2010)\citenamefont
  {Bendele}, \citenamefont {Amato}, \citenamefont {Conder}, \citenamefont
  {Elender}, \citenamefont {Keller}, \citenamefont {Klauss}, \citenamefont
  {Luetkens}, \citenamefont {Pomjakushina}, \citenamefont {Raselli},\ and\
  \citenamefont {Khasanov}}]{Bendele2010}%
  \BibitemOpen
  \bibfield  {author} {\bibinfo {author} {\bibfnamefont {M.}~\bibnamefont
  {Bendele}}, \bibinfo {author} {\bibfnamefont {A.}~\bibnamefont {Amato}},
  \bibinfo {author} {\bibfnamefont {K.}~\bibnamefont {Conder}}, \bibinfo
  {author} {\bibfnamefont {M.}~\bibnamefont {Elender}}, \bibinfo {author}
  {\bibfnamefont {H.}~\bibnamefont {Keller}}, \bibinfo {author} {\bibfnamefont
  {H.-H.}\ \bibnamefont {Klauss}}, \bibinfo {author} {\bibfnamefont
  {H.}~\bibnamefont {Luetkens}}, \bibinfo {author} {\bibfnamefont
  {E.}~\bibnamefont {Pomjakushina}}, \bibinfo {author} {\bibfnamefont
  {A.}~\bibnamefont {Raselli}}, \ and\ \bibinfo {author} {\bibfnamefont
  {R.}~\bibnamefont {Khasanov}},\ }\href {\doibase
  10.1103/PhysRevLett.104.087003} {\bibfield  {journal} {\bibinfo  {journal}
  {Phys. Rev. Lett.}\ }\textbf {\bibinfo {volume} {104}},\ \bibinfo {pages}
  {087003} (\bibinfo {year} {2010})}\BibitemShut {NoStop}%
\bibitem [{\citenamefont {Wang}\ \emph
  {et~al.}(2016{\natexlab{a}})\citenamefont {Wang}, \citenamefont {Shen},
  \citenamefont {Pan}, \citenamefont {Zhang}, \citenamefont {Ikeuchi},
  \citenamefont {Iida}, \citenamefont {Christianson}, \citenamefont {Walker},
  \citenamefont {Adroja}, \citenamefont {Abdel-Hafiez}, \citenamefont {Chen},
  \citenamefont {Chareev}, \citenamefont {Vasiliev},\ and\ \citenamefont
  {Zhao}}]{Wang2015GS}%
  \BibitemOpen
  \bibfield  {author} {\bibinfo {author} {\bibfnamefont {Q.}~\bibnamefont
  {Wang}}, \bibinfo {author} {\bibfnamefont {Y.}~\bibnamefont {Shen}}, \bibinfo
  {author} {\bibfnamefont {B.}~\bibnamefont {Pan}}, \bibinfo {author}
  {\bibfnamefont {X.}~\bibnamefont {Zhang}}, \bibinfo {author} {\bibfnamefont
  {K.}~\bibnamefont {Ikeuchi}}, \bibinfo {author} {\bibfnamefont
  {K.}~\bibnamefont {Iida}}, \bibinfo {author} {\bibfnamefont {A.~D.}\
  \bibnamefont {Christianson}}, \bibinfo {author} {\bibfnamefont {H.~C.}\
  \bibnamefont {Walker}}, \bibinfo {author} {\bibfnamefont {D.~T.}\
  \bibnamefont {Adroja}}, \bibinfo {author} {\bibfnamefont {M.}~\bibnamefont
  {Abdel-Hafiez}}, \bibinfo {author} {\bibfnamefont {X.}~\bibnamefont {Chen}},
  \bibinfo {author} {\bibfnamefont {D.~A.}\ \bibnamefont {Chareev}}, \bibinfo
  {author} {\bibfnamefont {A.~N.}\ \bibnamefont {Vasiliev}}, \ and\ \bibinfo
  {author} {\bibfnamefont {J.}~\bibnamefont {Zhao}},\ }\href
  {http://dx.doi.org/10.1038/ncomms12182} {\bibfield  {journal} {\bibinfo
  {journal} {Nat. Comm.}\ }\textbf {\bibinfo {volume} {7}},\ \bibinfo {pages}
  {12182} (\bibinfo {year} {2016}{\natexlab{a}})}\BibitemShut {NoStop}%
\bibitem [{\citenamefont {Hsu}\ \emph {et~al.}(2008)\citenamefont {Hsu},
  \citenamefont {Luo}, \citenamefont {Yeh}, \citenamefont {Chen}, \citenamefont
  {Huang}, \citenamefont {Wu}, \citenamefont {Lee}, \citenamefont {Huang},
  \citenamefont {Chu}, \citenamefont {Yan},\ and\ \citenamefont
  {Wu}}]{Hsu2008}%
  \BibitemOpen
  \bibfield  {author} {\bibinfo {author} {\bibfnamefont {F.-C.}\ \bibnamefont
  {Hsu}}, \bibinfo {author} {\bibfnamefont {J.-Y.}\ \bibnamefont {Luo}},
  \bibinfo {author} {\bibfnamefont {K.-W.}\ \bibnamefont {Yeh}}, \bibinfo
  {author} {\bibfnamefont {T.-K.}\ \bibnamefont {Chen}}, \bibinfo {author}
  {\bibfnamefont {T.-W.}\ \bibnamefont {Huang}}, \bibinfo {author}
  {\bibfnamefont {P.~M.}\ \bibnamefont {Wu}}, \bibinfo {author} {\bibfnamefont
  {Y.-C.}\ \bibnamefont {Lee}}, \bibinfo {author} {\bibfnamefont {Y.-L.}\
  \bibnamefont {Huang}}, \bibinfo {author} {\bibfnamefont {Y.-Y.}\ \bibnamefont
  {Chu}}, \bibinfo {author} {\bibfnamefont {D.-C.}\ \bibnamefont {Yan}}, \ and\
  \bibinfo {author} {\bibfnamefont {M.-K.}\ \bibnamefont {Wu}},\ }\href
  {\doibase 10.1073/pnas.0807325105} {\bibfield  {journal} {\bibinfo  {journal}
  {Proceedings of the National Academy of Sciences}\ }\textbf {\bibinfo
  {volume} {105}},\ \bibinfo {pages} {14262} (\bibinfo {year}
  {2008})}\BibitemShut {NoStop}%
\bibitem [{\citenamefont {Mizuguchi}\ \emph {et~al.}(2008)\citenamefont
  {Mizuguchi}, \citenamefont {Tomioka}, \citenamefont {Tsuda}, \citenamefont
  {Yamaguchi},\ and\ \citenamefont {Takano}}]{Mizuguchi2008}%
  \BibitemOpen
  \bibfield  {author} {\bibinfo {author} {\bibfnamefont {Y.}~\bibnamefont
  {Mizuguchi}}, \bibinfo {author} {\bibfnamefont {F.}~\bibnamefont {Tomioka}},
  \bibinfo {author} {\bibfnamefont {S.}~\bibnamefont {Tsuda}}, \bibinfo
  {author} {\bibfnamefont {T.}~\bibnamefont {Yamaguchi}}, \ and\ \bibinfo
  {author} {\bibfnamefont {Y.}~\bibnamefont {Takano}},\ }\href {\doibase
  http://dx.doi.org/10.1063/1.3000616} {\bibfield  {journal} {\bibinfo
  {journal} {Applied Physics Letters}\ }\textbf {\bibinfo {volume} {93}},\
  \bibinfo {pages} {152505} (\bibinfo {year} {2008})}\BibitemShut {NoStop}%
\bibitem [{\citenamefont {Medvedev}\ \emph {et~al.}(2009)\citenamefont
  {Medvedev}, \citenamefont {McQueen}, \citenamefont {Troyan}, \citenamefont
  {Palasyuk}, \citenamefont {Eremets}, \citenamefont {Cava}, \citenamefont
  {Naghavi}, \citenamefont {Casper}, \citenamefont {Ksenofontov}, \citenamefont
  {Wortmann},\ and\ \citenamefont {Felser}}]{Medvedev2009}%
  \BibitemOpen
  \bibfield  {author} {\bibinfo {author} {\bibfnamefont {S.}~\bibnamefont
  {Medvedev}}, \bibinfo {author} {\bibfnamefont {T.~M.}\ \bibnamefont
  {McQueen}}, \bibinfo {author} {\bibfnamefont {I.~A.}\ \bibnamefont {Troyan}},
  \bibinfo {author} {\bibfnamefont {T.}~\bibnamefont {Palasyuk}}, \bibinfo
  {author} {\bibfnamefont {M.~I.}\ \bibnamefont {Eremets}}, \bibinfo {author}
  {\bibfnamefont {R.~J.}\ \bibnamefont {Cava}}, \bibinfo {author}
  {\bibfnamefont {S.}~\bibnamefont {Naghavi}}, \bibinfo {author} {\bibfnamefont
  {F.}~\bibnamefont {Casper}}, \bibinfo {author} {\bibfnamefont
  {V.}~\bibnamefont {Ksenofontov}}, \bibinfo {author} {\bibfnamefont
  {G.}~\bibnamefont {Wortmann}}, \ and\ \bibinfo {author} {\bibfnamefont
  {C.}~\bibnamefont {Felser}},\ }\href {\doibase 10.1038/nmat2491} {\bibfield
  {journal} {\bibinfo  {journal} {Nature Mat.}\ }\textbf {\bibinfo {volume}
  {8}},\ \bibinfo {pages} {630} (\bibinfo {year} {2009})}\BibitemShut {NoStop}%
\bibitem [{\citenamefont {Margadonna}\ \emph {et~al.}(2009)\citenamefont
  {Margadonna}, \citenamefont {Takabayashi}, \citenamefont {Ohishi},
  \citenamefont {Mizuguchi}, \citenamefont {Takano}, \citenamefont {Kagayama},
  \citenamefont {Nakagawa}, \citenamefont {Takata},\ and\ \citenamefont
  {Prassides}}]{Margadonna2009}%
  \BibitemOpen
  \bibfield  {author} {\bibinfo {author} {\bibfnamefont {S.}~\bibnamefont
  {Margadonna}}, \bibinfo {author} {\bibfnamefont {Y.}~\bibnamefont
  {Takabayashi}}, \bibinfo {author} {\bibfnamefont {Y.}~\bibnamefont {Ohishi}},
  \bibinfo {author} {\bibfnamefont {Y.}~\bibnamefont {Mizuguchi}}, \bibinfo
  {author} {\bibfnamefont {Y.}~\bibnamefont {Takano}}, \bibinfo {author}
  {\bibfnamefont {T.}~\bibnamefont {Kagayama}}, \bibinfo {author}
  {\bibfnamefont {T.}~\bibnamefont {Nakagawa}}, \bibinfo {author}
  {\bibfnamefont {M.}~\bibnamefont {Takata}}, \ and\ \bibinfo {author}
  {\bibfnamefont {K.}~\bibnamefont {Prassides}},\ }\href {\doibase
  10.1103/PhysRevB.80.064506} {\bibfield  {journal} {\bibinfo  {journal} {Phys.
  Rev. B}\ }\textbf {\bibinfo {volume} {80}},\ \bibinfo {pages} {064506}
  (\bibinfo {year} {2009})}\BibitemShut {NoStop}%
\bibitem [{\citenamefont {Garbarino}\ \emph {et~al.}(2009)\citenamefont
  {Garbarino}, \citenamefont {Sow}, \citenamefont {Lejay}, \citenamefont
  {Sulpice}, \citenamefont {Toulemonde}, \citenamefont {Mezouar},\ and\
  \citenamefont {N{\'u}{\~n}ez-Regueiro}}]{Garbarino2009}%
  \BibitemOpen
  \bibfield  {author} {\bibinfo {author} {\bibfnamefont {G.}~\bibnamefont
  {Garbarino}}, \bibinfo {author} {\bibfnamefont {A.}~\bibnamefont {Sow}},
  \bibinfo {author} {\bibfnamefont {P.}~\bibnamefont {Lejay}}, \bibinfo
  {author} {\bibfnamefont {A.}~\bibnamefont {Sulpice}}, \bibinfo {author}
  {\bibfnamefont {P.}~\bibnamefont {Toulemonde}}, \bibinfo {author}
  {\bibfnamefont {M.}~\bibnamefont {Mezouar}}, \ and\ \bibinfo {author}
  {\bibfnamefont {M.}~\bibnamefont {N{\'u}{\~n}ez-Regueiro}},\ }\href
  {http://stacks.iop.org/0295-5075/86/i=2/a=27001} {\bibfield  {journal}
  {\bibinfo  {journal} {EPL (Europhysics Letters)}\ }\textbf {\bibinfo {volume}
  {86}},\ \bibinfo {pages} {27001} (\bibinfo {year} {2009})}\BibitemShut
  {NoStop}%
\bibitem [{\citenamefont {Ge}\ \emph {et~al.}(2015)\citenamefont {Ge},
  \citenamefont {Liu}, \citenamefont {Liu}, \citenamefont {Gao}, \citenamefont
  {Qian}, \citenamefont {Xue}, \citenamefont {Liu},\ and\ \citenamefont
  {Jia}}]{Ge2015}%
  \BibitemOpen
  \bibfield  {author} {\bibinfo {author} {\bibfnamefont {J.-F.}\ \bibnamefont
  {Ge}}, \bibinfo {author} {\bibfnamefont {Z.-L.}\ \bibnamefont {Liu}},
  \bibinfo {author} {\bibfnamefont {C.}~\bibnamefont {Liu}}, \bibinfo {author}
  {\bibfnamefont {C.-L.}\ \bibnamefont {Gao}}, \bibinfo {author} {\bibfnamefont
  {D.}~\bibnamefont {Qian}}, \bibinfo {author} {\bibfnamefont {Q.-K.}\
  \bibnamefont {Xue}}, \bibinfo {author} {\bibfnamefont {Y.}~\bibnamefont
  {Liu}}, \ and\ \bibinfo {author} {\bibfnamefont {J.-F.}\ \bibnamefont
  {Jia}},\ }\href {\doibase 10.1038/nmat4153} {\bibfield  {journal} {\bibinfo
  {journal} {Nature Materials}\ }\textbf {\bibinfo {volume} {14}},\ \bibinfo
  {pages} {285} (\bibinfo {year} {2015})}\BibitemShut {NoStop}%
\bibitem [{\citenamefont {Kaluarachchi}\ \emph {et~al.}(2016)\citenamefont
  {Kaluarachchi}, \citenamefont {Taufour}, \citenamefont {B\"ohmer},
  \citenamefont {Tanatar}, \citenamefont {Bud{'}ko}, \citenamefont {Kogan},
  \citenamefont {Prozorov},\ and\ \citenamefont {Canfield}}]{Kaluarachchi2016}%
  \BibitemOpen
  \bibfield  {author} {\bibinfo {author} {\bibfnamefont {U.~S.}\ \bibnamefont
  {Kaluarachchi}}, \bibinfo {author} {\bibfnamefont {V.}~\bibnamefont
  {Taufour}}, \bibinfo {author} {\bibfnamefont {A.~E.}\ \bibnamefont
  {B\"ohmer}}, \bibinfo {author} {\bibfnamefont {M.~A.}\ \bibnamefont
  {Tanatar}}, \bibinfo {author} {\bibfnamefont {S.~L.}\ \bibnamefont
  {Bud{'}ko}}, \bibinfo {author} {\bibfnamefont {V.~G.}\ \bibnamefont {Kogan}},
  \bibinfo {author} {\bibfnamefont {R.}~\bibnamefont {Prozorov}}, \ and\
  \bibinfo {author} {\bibfnamefont {P.~C.}\ \bibnamefont {Canfield}},\ }\href
  {\doibase 10.1103/PhysRevB.93.064503} {\bibfield  {journal} {\bibinfo
  {journal} {Phys. Rev. B}\ }\textbf {\bibinfo {volume} {93}},\ \bibinfo
  {pages} {064503} (\bibinfo {year} {2016})}\BibitemShut {NoStop}%
\bibitem [{\citenamefont {Sun}\ \emph {et~al.}(2016)\citenamefont {Sun},
  \citenamefont {Matsuura}, \citenamefont {Ye}, \citenamefont {Mizukami},
  \citenamefont {Shimozawa}, \citenamefont {Matsubayashi}, \citenamefont
  {Yamashita}, \citenamefont {Watashige}, \citenamefont {Kasahara},
  \citenamefont {Matsuda}, \citenamefont {Yan}, \citenamefont {Sales},
  \citenamefont {Uwatoko}, \citenamefont {Cheng},\ and\ \citenamefont
  {Shibauchi}}]{Sun_2016}%
  \BibitemOpen
  \bibfield  {author} {\bibinfo {author} {\bibfnamefont {J.~P.}\ \bibnamefont
  {Sun}}, \bibinfo {author} {\bibfnamefont {K.}~\bibnamefont {Matsuura}},
  \bibinfo {author} {\bibfnamefont {G.~Z.}\ \bibnamefont {Ye}}, \bibinfo
  {author} {\bibfnamefont {Y.}~\bibnamefont {Mizukami}}, \bibinfo {author}
  {\bibfnamefont {M.}~\bibnamefont {Shimozawa}}, \bibinfo {author}
  {\bibfnamefont {K.}~\bibnamefont {Matsubayashi}}, \bibinfo {author}
  {\bibfnamefont {M.}~\bibnamefont {Yamashita}}, \bibinfo {author}
  {\bibfnamefont {T.}~\bibnamefont {Watashige}}, \bibinfo {author}
  {\bibfnamefont {S.}~\bibnamefont {Kasahara}}, \bibinfo {author}
  {\bibfnamefont {Y.}~\bibnamefont {Matsuda}}, \bibinfo {author} {\bibfnamefont
  {J.~Q.}\ \bibnamefont {Yan}}, \bibinfo {author} {\bibfnamefont {B.~C.}\
  \bibnamefont {Sales}}, \bibinfo {author} {\bibfnamefont {Y.}~\bibnamefont
  {Uwatoko}}, \bibinfo {author} {\bibfnamefont {J.~G.}\ \bibnamefont {Cheng}},
  \ and\ \bibinfo {author} {\bibfnamefont {T.}~\bibnamefont {Shibauchi}},\
  }\href {http://dx.doi.org/10.1038/ncomms12146} {\bibfield  {journal}
  {\bibinfo  {journal} {Nature Communications}\ }\textbf {\bibinfo {volume}
  {7}},\ \bibinfo {pages} {12146} (\bibinfo {year} {2016})}\BibitemShut
  {NoStop}%
\bibitem [{\citenamefont {Miyoshi}\ \emph {et~al.}(2014)\citenamefont
  {Miyoshi}, \citenamefont {Morishita}, \citenamefont {Mutou}, \citenamefont
  {Kondo}, \citenamefont {Seida}, \citenamefont {Fujiwara}, \citenamefont
  {Takeuchi},\ and\ \citenamefont {Nishigori}}]{Miyoshi2014}%
  \BibitemOpen
  \bibfield  {author} {\bibinfo {author} {\bibfnamefont {K.}~\bibnamefont
  {Miyoshi}}, \bibinfo {author} {\bibfnamefont {K.}~\bibnamefont {Morishita}},
  \bibinfo {author} {\bibfnamefont {E.}~\bibnamefont {Mutou}}, \bibinfo
  {author} {\bibfnamefont {M.}~\bibnamefont {Kondo}}, \bibinfo {author}
  {\bibfnamefont {O.}~\bibnamefont {Seida}}, \bibinfo {author} {\bibfnamefont
  {K.}~\bibnamefont {Fujiwara}}, \bibinfo {author} {\bibfnamefont
  {J.}~\bibnamefont {Takeuchi}}, \ and\ \bibinfo {author} {\bibfnamefont
  {S.}~\bibnamefont {Nishigori}},\ }\href {\doibase 10.7566/JPSJ.83.013702}
  {\bibfield  {journal} {\bibinfo  {journal} {Journal of the Physical Society
  of Japan}\ }\textbf {\bibinfo {volume} {83}},\ \bibinfo {pages} {013702}
  (\bibinfo {year} {2014})}\BibitemShut {NoStop}%
\bibitem [{\citenamefont {Terashima}\ \emph {et~al.}(2015)\citenamefont
  {Terashima}, \citenamefont {Kikugawa}, \citenamefont {Kasahara},
  \citenamefont {Watashige}, \citenamefont {Shibauchi}, \citenamefont
  {Matsuda}, \citenamefont {Wolf}, \citenamefont {B\"ohmer}, \citenamefont
  {Hardy}, \citenamefont {Meingast}, \citenamefont {v.~L\"ohneysen},\ and\
  \citenamefont {Uji}}]{Terashima2015}%
  \BibitemOpen
  \bibfield  {author} {\bibinfo {author} {\bibfnamefont {T.}~\bibnamefont
  {Terashima}}, \bibinfo {author} {\bibfnamefont {N.}~\bibnamefont {Kikugawa}},
  \bibinfo {author} {\bibfnamefont {S.}~\bibnamefont {Kasahara}}, \bibinfo
  {author} {\bibfnamefont {T.}~\bibnamefont {Watashige}}, \bibinfo {author}
  {\bibfnamefont {T.}~\bibnamefont {Shibauchi}}, \bibinfo {author}
  {\bibfnamefont {Y.}~\bibnamefont {Matsuda}}, \bibinfo {author} {\bibfnamefont
  {T.}~\bibnamefont {Wolf}}, \bibinfo {author} {\bibfnamefont {A.~E.}\
  \bibnamefont {B\"ohmer}}, \bibinfo {author} {\bibfnamefont {F.}~\bibnamefont
  {Hardy}}, \bibinfo {author} {\bibfnamefont {C.}~\bibnamefont {Meingast}},
  \bibinfo {author} {\bibfnamefont {H.}~\bibnamefont {v.~L\"ohneysen}}, \ and\
  \bibinfo {author} {\bibfnamefont {S.}~\bibnamefont {Uji}},\ }\href {\doibase
  10.7566/JPSJ.84.063701} {\bibfield  {journal} {\bibinfo  {journal} {Journal
  of the Physical Society of Japan}\ }\textbf {\bibinfo {volume} {84}},\
  \bibinfo {pages} {063701} (\bibinfo {year} {2015})}\BibitemShut {NoStop}%
\bibitem [{\citenamefont {{Kothapalli}}\ \emph {et~al.}(2016)\citenamefont
  {{Kothapalli}}, \citenamefont {{B{\"o}hmer}}, \citenamefont {{Jayasekara}},
  \citenamefont {{Ueland}}, \citenamefont {{Das}}, \citenamefont {{Sapkota}},
  \citenamefont {{Taufour}}, \citenamefont {{Xiao}}, \citenamefont {{Alp}},
  \citenamefont {{Bud'ko}}, \citenamefont {{Canfield}}, \citenamefont
  {{Kreyssig}},\ and\ \citenamefont {{Goldman}}}]{Kothapalli2016}%
  \BibitemOpen
  \bibfield  {author} {\bibinfo {author} {\bibfnamefont {K.}~\bibnamefont
  {{Kothapalli}}}, \bibinfo {author} {\bibfnamefont {A.~E.}\ \bibnamefont
  {{B{\"o}hmer}}}, \bibinfo {author} {\bibfnamefont {W.~T.}\ \bibnamefont
  {{Jayasekara}}}, \bibinfo {author} {\bibfnamefont {B.~G.}\ \bibnamefont
  {{Ueland}}}, \bibinfo {author} {\bibfnamefont {P.}~\bibnamefont {{Das}}},
  \bibinfo {author} {\bibfnamefont {A.}~\bibnamefont {{Sapkota}}}, \bibinfo
  {author} {\bibfnamefont {V.}~\bibnamefont {{Taufour}}}, \bibinfo {author}
  {\bibfnamefont {Y.}~\bibnamefont {{Xiao}}}, \bibinfo {author} {\bibfnamefont
  {E.~E.}\ \bibnamefont {{Alp}}}, \bibinfo {author} {\bibfnamefont {S.~L.}\
  \bibnamefont {{Bud'ko}}}, \bibinfo {author} {\bibfnamefont {P.~C.}\
  \bibnamefont {{Canfield}}}, \bibinfo {author} {\bibfnamefont
  {A.}~\bibnamefont {{Kreyssig}}}, \ and\ \bibinfo {author} {\bibfnamefont
  {A.~I.}\ \bibnamefont {{Goldman}}},\ }\href
  {http://dx.doi.org/10.1038/ncomms12728} {\bibfield  {journal} {\bibinfo
  {journal} {Nature Communications}\ }\textbf {\bibinfo {volume} {7}},\
  \bibinfo {pages} {12728} (\bibinfo {year} {2016})}\BibitemShut {NoStop}%
\bibitem [{\citenamefont {Wang}\ \emph
  {et~al.}(2016{\natexlab{b}})\citenamefont {Wang}, \citenamefont {Sun},
  \citenamefont {Cui}, \citenamefont {Song}, \citenamefont {Li}, \citenamefont
  {Yu}, \citenamefont {Lei},\ and\ \citenamefont {Yu}}]{Wang2016}%
  \BibitemOpen
  \bibfield  {author} {\bibinfo {author} {\bibfnamefont {P.~S.}\ \bibnamefont
  {Wang}}, \bibinfo {author} {\bibfnamefont {S.~S.}\ \bibnamefont {Sun}},
  \bibinfo {author} {\bibfnamefont {Y.}~\bibnamefont {Cui}}, \bibinfo {author}
  {\bibfnamefont {W.~H.}\ \bibnamefont {Song}}, \bibinfo {author}
  {\bibfnamefont {T.~R.}\ \bibnamefont {Li}}, \bibinfo {author} {\bibfnamefont
  {R.}~\bibnamefont {Yu}}, \bibinfo {author} {\bibfnamefont {H.}~\bibnamefont
  {Lei}}, \ and\ \bibinfo {author} {\bibfnamefont {W.}~\bibnamefont {Yu}},\
  }\href {\doibase 10.1103/PhysRevLett.117.237001} {\bibfield  {journal}
  {\bibinfo  {journal} {Phys. Rev. Lett.}\ }\textbf {\bibinfo {volume} {117}},\
  \bibinfo {pages} {237001} (\bibinfo {year} {2016}{\natexlab{b}})}\BibitemShut
  {NoStop}%
\bibitem [{\citenamefont {Bendele}\ \emph {et~al.}(2012)\citenamefont
  {Bendele}, \citenamefont {Ichsanow}, \citenamefont {Pashkevich},
  \citenamefont {Keller}, \citenamefont {Str\"assle}, \citenamefont {Gusev},
  \citenamefont {Pomjakushina}, \citenamefont {Conder}, \citenamefont
  {Khasanov},\ and\ \citenamefont {Keller}}]{Bendele2012}%
  \BibitemOpen
  \bibfield  {author} {\bibinfo {author} {\bibfnamefont {M.}~\bibnamefont
  {Bendele}}, \bibinfo {author} {\bibfnamefont {A.}~\bibnamefont {Ichsanow}},
  \bibinfo {author} {\bibfnamefont {Y.}~\bibnamefont {Pashkevich}}, \bibinfo
  {author} {\bibfnamefont {L.}~\bibnamefont {Keller}}, \bibinfo {author}
  {\bibfnamefont {T.}~\bibnamefont {Str\"assle}}, \bibinfo {author}
  {\bibfnamefont {A.}~\bibnamefont {Gusev}}, \bibinfo {author} {\bibfnamefont
  {E.}~\bibnamefont {Pomjakushina}}, \bibinfo {author} {\bibfnamefont
  {K.}~\bibnamefont {Conder}}, \bibinfo {author} {\bibfnamefont
  {R.}~\bibnamefont {Khasanov}}, \ and\ \bibinfo {author} {\bibfnamefont
  {H.}~\bibnamefont {Keller}},\ }\href {\doibase 10.1103/PhysRevB.85.064517}
  {\bibfield  {journal} {\bibinfo  {journal} {Phys. Rev. B}\ }\textbf {\bibinfo
  {volume} {85}},\ \bibinfo {pages} {064517} (\bibinfo {year}
  {2012})}\BibitemShut {NoStop}%
\bibitem [{\citenamefont {B{\"{u}}chner}\ and\ \citenamefont
  {Hess}(2009)}]{Buechner2009}%
  \BibitemOpen
  \bibfield  {author} {\bibinfo {author} {\bibfnamefont {B.}~\bibnamefont
  {B{\"{u}}chner}}\ and\ \bibinfo {author} {\bibfnamefont {C.}~\bibnamefont
  {Hess}},\ }\href {\doibase 10.1038/nmat2501} {\bibfield  {journal} {\bibinfo
  {journal} {Nature Materials}\ }\textbf {\bibinfo {volume} {8}},\ \bibinfo
  {pages} {615} (\bibinfo {year} {2009})}\BibitemShut {NoStop}%
\bibitem [{\citenamefont {Khasanov}\ \emph {et~al.}(2017)\citenamefont
  {Khasanov}, \citenamefont {Guguchia}, \citenamefont {Amato}, \citenamefont
  {Morenzoni}, \citenamefont {Dong}, \citenamefont {Zhou},\ and\ \citenamefont
  {Zhao}}]{Khasanov2016}%
  \BibitemOpen
  \bibfield  {author} {\bibinfo {author} {\bibfnamefont {R.}~\bibnamefont
  {Khasanov}}, \bibinfo {author} {\bibfnamefont {Z.}~\bibnamefont {Guguchia}},
  \bibinfo {author} {\bibfnamefont {A.}~\bibnamefont {Amato}}, \bibinfo
  {author} {\bibfnamefont {E.}~\bibnamefont {Morenzoni}}, \bibinfo {author}
  {\bibfnamefont {X.}~\bibnamefont {Dong}}, \bibinfo {author} {\bibfnamefont
  {F.}~\bibnamefont {Zhou}}, \ and\ \bibinfo {author} {\bibfnamefont
  {Z.}~\bibnamefont {Zhao}},\ }\href {\doibase 10.1103/PhysRevB.95.180504}
  {\bibfield  {journal} {\bibinfo  {journal} {Phys. Rev. B}\ }\textbf {\bibinfo
  {volume} {95}},\ \bibinfo {pages} {180504} (\bibinfo {year}
  {2017})}\BibitemShut {NoStop}%
\bibitem [{\citenamefont {Kumar}\ \emph {et~al.}(2010)\citenamefont {Kumar},
  \citenamefont {Zhang}, \citenamefont {Sinogeikin}, \citenamefont {Xiao},
  \citenamefont {Kumar}, \citenamefont {Chow}, \citenamefont {Cornelius},\ and\
  \citenamefont {Chen}}]{Kumar2010}%
  \BibitemOpen
  \bibfield  {author} {\bibinfo {author} {\bibfnamefont {R.~S.}\ \bibnamefont
  {Kumar}}, \bibinfo {author} {\bibfnamefont {Y.}~\bibnamefont {Zhang}},
  \bibinfo {author} {\bibfnamefont {S.}~\bibnamefont {Sinogeikin}}, \bibinfo
  {author} {\bibfnamefont {Y.}~\bibnamefont {Xiao}}, \bibinfo {author}
  {\bibfnamefont {S.}~\bibnamefont {Kumar}}, \bibinfo {author} {\bibfnamefont
  {P.}~\bibnamefont {Chow}}, \bibinfo {author} {\bibfnamefont {A.~L.}\
  \bibnamefont {Cornelius}}, \ and\ \bibinfo {author} {\bibfnamefont
  {C.}~\bibnamefont {Chen}},\ }\href {\doibase 10.1021/jp1060446} {\bibfield
  {journal} {\bibinfo  {journal} {The Journal of Physical Chemistry B}\
  }\textbf {\bibinfo {volume} {114}},\ \bibinfo {pages} {12597} (\bibinfo
  {year} {2010})}\BibitemShut {NoStop}%
\bibitem [{\citenamefont {Svitlyk}\ \emph {et~al.}(2017)\citenamefont
  {Svitlyk}, \citenamefont {Raba}, \citenamefont {Dmitriev}, \citenamefont
  {Rodi\`ere}, \citenamefont {Toulemonde}, \citenamefont {Chernyshov},\ and\
  \citenamefont {Mezouar}}]{Svitlyk2016}%
  \BibitemOpen
  \bibfield  {author} {\bibinfo {author} {\bibfnamefont {V.}~\bibnamefont
  {Svitlyk}}, \bibinfo {author} {\bibfnamefont {M.}~\bibnamefont {Raba}},
  \bibinfo {author} {\bibfnamefont {V.}~\bibnamefont {Dmitriev}}, \bibinfo
  {author} {\bibfnamefont {P.}~\bibnamefont {Rodi\`ere}}, \bibinfo {author}
  {\bibfnamefont {P.}~\bibnamefont {Toulemonde}}, \bibinfo {author}
  {\bibfnamefont {D.}~\bibnamefont {Chernyshov}}, \ and\ \bibinfo {author}
  {\bibfnamefont {M.}~\bibnamefont {Mezouar}},\ }\href {\doibase
  10.1103/PhysRevB.96.014520} {\bibfield  {journal} {\bibinfo  {journal} {Phys.
  Rev. B}\ }\textbf {\bibinfo {volume} {96}},\ \bibinfo {pages} {014520}
  (\bibinfo {year} {2017})}\BibitemShut {NoStop}%
\bibitem [{\citenamefont {B{\"o}hmer}\ \emph {et~al.}(2016)\citenamefont
  {B{\"o}hmer}, \citenamefont {Taufour}, \citenamefont {Straszheim},
  \citenamefont {Wolf},\ and\ \citenamefont {Canfield}}]{Boehmer2016II}%
  \BibitemOpen
  \bibfield  {author} {\bibinfo {author} {\bibfnamefont {A.~E.}\ \bibnamefont
  {B{\"o}hmer}}, \bibinfo {author} {\bibfnamefont {V.}~\bibnamefont {Taufour}},
  \bibinfo {author} {\bibfnamefont {W.~E.}\ \bibnamefont {Straszheim}},
  \bibinfo {author} {\bibfnamefont {T.}~\bibnamefont {Wolf}}, \ and\ \bibinfo
  {author} {\bibfnamefont {P.~C.}\ \bibnamefont {Canfield}},\ }\href {\doibase
  10.1103/PhysRevB.94.024526} {\bibfield  {journal} {\bibinfo  {journal} {Phys.
  Rev. B}\ }\textbf {\bibinfo {volume} {94}},\ \bibinfo {pages} {024526}
  (\bibinfo {year} {2016})}\BibitemShut {NoStop}%
\bibitem [{\citenamefont {Bi}\ \emph {et~al.}(2015)\citenamefont {Bi},
  \citenamefont {Zhao}, \citenamefont {Lin}, \citenamefont {Jia}, \citenamefont
  {Hu}, \citenamefont {Jin}, \citenamefont {Ferry}, \citenamefont {Yang},
  \citenamefont {Struzhkin},\ and\ \citenamefont {Alp}}]{Bi2015}%
  \BibitemOpen
  \bibfield  {author} {\bibinfo {author} {\bibfnamefont {W.}~\bibnamefont
  {Bi}}, \bibinfo {author} {\bibfnamefont {J.}~\bibnamefont {Zhao}}, \bibinfo
  {author} {\bibfnamefont {J.}~\bibnamefont {Lin}}, \bibinfo {author}
  {\bibfnamefont {Q.}~\bibnamefont {Jia}}, \bibinfo {author} {\bibfnamefont
  {M.~Y.}\ \bibnamefont {Hu}}, \bibinfo {author} {\bibfnamefont
  {C.}~\bibnamefont {Jin}}, \bibinfo {author} {\bibfnamefont {R.}~\bibnamefont
  {Ferry}}, \bibinfo {author} {\bibfnamefont {W.}~\bibnamefont {Yang}},
  \bibinfo {author} {\bibfnamefont {V.}~\bibnamefont {Struzhkin}}, \ and\
  \bibinfo {author} {\bibfnamefont {E.~E.}\ \bibnamefont {Alp}},\ }\href
  {\doibase 10.1107/S1600577515003586} {\bibfield  {journal} {\bibinfo
  {journal} {Journal of Synchrotron Radiation}\ }\textbf {\bibinfo {volume}
  {22}},\ \bibinfo {pages} {760} (\bibinfo {year} {2015})}\BibitemShut
  {NoStop}%
\bibitem [{\citenamefont {Sturhahn}(2000)}]{Sturhahn2000}%
  \BibitemOpen
  \bibfield  {author} {\bibinfo {author} {\bibfnamefont {W.}~\bibnamefont
  {Sturhahn}},\ }\href@noop {} {\bibfield  {journal} {\bibinfo  {journal}
  {Hyperfine Interact.}\ }\textbf {\bibinfo {volume} {125}},\ \bibinfo {pages}
  {149–172} (\bibinfo {year} {2000})}\BibitemShut {NoStop}%
\bibitem [{\citenamefont {B\"ohmer}\ \emph {et~al.}(2015)\citenamefont
  {B\"ohmer}, \citenamefont {Arai}, \citenamefont {Hardy}, \citenamefont
  {Hattori}, \citenamefont {Iye}, \citenamefont {Wolf}, \citenamefont
  {L\"ohneysen}, \citenamefont {Ishida},\ and\ \citenamefont
  {Meingast}}]{Boehmer2015}%
  \BibitemOpen
  \bibfield  {author} {\bibinfo {author} {\bibfnamefont {A.~E.}\ \bibnamefont
  {B\"ohmer}}, \bibinfo {author} {\bibfnamefont {T.}~\bibnamefont {Arai}},
  \bibinfo {author} {\bibfnamefont {F.}~\bibnamefont {Hardy}}, \bibinfo
  {author} {\bibfnamefont {T.}~\bibnamefont {Hattori}}, \bibinfo {author}
  {\bibfnamefont {T.}~\bibnamefont {Iye}}, \bibinfo {author} {\bibfnamefont
  {T.}~\bibnamefont {Wolf}}, \bibinfo {author} {\bibfnamefont {H.~v.}\
  \bibnamefont {L\"ohneysen}}, \bibinfo {author} {\bibfnamefont
  {K.}~\bibnamefont {Ishida}}, \ and\ \bibinfo {author} {\bibfnamefont
  {C.}~\bibnamefont {Meingast}},\ }\href {\doibase
  10.1103/PhysRevLett.114.027001} {\bibfield  {journal} {\bibinfo  {journal}
  {Phys. Rev. Lett.}\ }\textbf {\bibinfo {volume} {114}},\ \bibinfo {pages}
  {027001} (\bibinfo {year} {2015})}\BibitemShut {NoStop}%
\bibitem [{\citenamefont {Meingast}\ \emph {et~al.}(2012)\citenamefont
  {Meingast}, \citenamefont {Hardy}, \citenamefont {Heid}, \citenamefont
  {Adelmann}, \citenamefont {B\"ohmer}, \citenamefont {Burger}, \citenamefont
  {Ernst}, \citenamefont {Fromknecht}, \citenamefont {Schweiss},\ and\
  \citenamefont {Wolf}}]{Meingast2012}%
  \BibitemOpen
  \bibfield  {author} {\bibinfo {author} {\bibfnamefont {C.}~\bibnamefont
  {Meingast}}, \bibinfo {author} {\bibfnamefont {F.}~\bibnamefont {Hardy}},
  \bibinfo {author} {\bibfnamefont {R.}~\bibnamefont {Heid}}, \bibinfo {author}
  {\bibfnamefont {P.}~\bibnamefont {Adelmann}}, \bibinfo {author}
  {\bibfnamefont {A.}~\bibnamefont {B\"ohmer}}, \bibinfo {author}
  {\bibfnamefont {P.}~\bibnamefont {Burger}}, \bibinfo {author} {\bibfnamefont
  {D.}~\bibnamefont {Ernst}}, \bibinfo {author} {\bibfnamefont
  {R.}~\bibnamefont {Fromknecht}}, \bibinfo {author} {\bibfnamefont
  {P.}~\bibnamefont {Schweiss}}, \ and\ \bibinfo {author} {\bibfnamefont
  {T.}~\bibnamefont {Wolf}},\ }\href {\doibase 10.1103/PhysRevLett.108.177004}
  {\bibfield  {journal} {\bibinfo  {journal} {Phys. Rev. Lett.}\ }\textbf
  {\bibinfo {volume} {108}},\ \bibinfo {pages} {177004} (\bibinfo {year}
  {2012})}\BibitemShut {NoStop}%
\bibitem [{\citenamefont {Hardy}\ \emph {et~al.}(2010)\citenamefont {Hardy},
  \citenamefont {Wolf}, \citenamefont {Fisher}, \citenamefont {Eder},
  \citenamefont {Schweiss}, \citenamefont {Adelmann}, \citenamefont
  {v.~L\"ohneysen},\ and\ \citenamefont {Meingast}}]{Hardy2010}%
  \BibitemOpen
  \bibfield  {author} {\bibinfo {author} {\bibfnamefont {F.}~\bibnamefont
  {Hardy}}, \bibinfo {author} {\bibfnamefont {T.}~\bibnamefont {Wolf}},
  \bibinfo {author} {\bibfnamefont {R.~A.}\ \bibnamefont {Fisher}}, \bibinfo
  {author} {\bibfnamefont {R.}~\bibnamefont {Eder}}, \bibinfo {author}
  {\bibfnamefont {P.}~\bibnamefont {Schweiss}}, \bibinfo {author}
  {\bibfnamefont {P.}~\bibnamefont {Adelmann}}, \bibinfo {author}
  {\bibfnamefont {H.}~\bibnamefont {v.~L\"ohneysen}}, \ and\ \bibinfo {author}
  {\bibfnamefont {C.}~\bibnamefont {Meingast}},\ }\href {\doibase
  10.1103/PhysRevB.81.060501} {\bibfield  {journal} {\bibinfo  {journal} {Phys.
  Rev. B}\ }\textbf {\bibinfo {volume} {81}},\ \bibinfo {pages} {060501}
  (\bibinfo {year} {2010})}\BibitemShut {NoStop}%
\bibitem [{\citenamefont {Prozorov}\ \emph {et~al.}(2009)\citenamefont
  {Prozorov}, \citenamefont {Tanatar}, \citenamefont {Ni}, \citenamefont
  {Kreyssig}, \citenamefont {Nandi}, \citenamefont {Bud'ko}, \citenamefont
  {Goldman},\ and\ \citenamefont {Canfield}}]{Prozorov2009}%
  \BibitemOpen
  \bibfield  {author} {\bibinfo {author} {\bibfnamefont {R.}~\bibnamefont
  {Prozorov}}, \bibinfo {author} {\bibfnamefont {M.~A.}\ \bibnamefont
  {Tanatar}}, \bibinfo {author} {\bibfnamefont {N.}~\bibnamefont {Ni}},
  \bibinfo {author} {\bibfnamefont {A.}~\bibnamefont {Kreyssig}}, \bibinfo
  {author} {\bibfnamefont {S.}~\bibnamefont {Nandi}}, \bibinfo {author}
  {\bibfnamefont {S.~L.}\ \bibnamefont {Bud'ko}}, \bibinfo {author}
  {\bibfnamefont {A.~I.}\ \bibnamefont {Goldman}}, \ and\ \bibinfo {author}
  {\bibfnamefont {P.~C.}\ \bibnamefont {Canfield}},\ }\href {\doibase
  10.1103/PhysRevB.80.174517} {\bibfield  {journal} {\bibinfo  {journal} {Phys.
  Rev. B}\ }\textbf {\bibinfo {volume} {80}},\ \bibinfo {pages} {174517}
  (\bibinfo {year} {2009})}\BibitemShut {NoStop}%
\bibitem [{\citenamefont {Avci}\ \emph {et~al.}(2012)\citenamefont {Avci},
  \citenamefont {Chmaissem}, \citenamefont {Chung}, \citenamefont {Rosenkranz},
  \citenamefont {Goremychkin}, \citenamefont {Castellan}, \citenamefont
  {Todorov}, \citenamefont {Schlueter}, \citenamefont {Claus}, \citenamefont
  {Daoud-Aladine}, \citenamefont {Khalyavin}, \citenamefont {Kanatzidis},\ and\
  \citenamefont {Osborn}}]{Avci2012}%
  \BibitemOpen
  \bibfield  {author} {\bibinfo {author} {\bibfnamefont {S.}~\bibnamefont
  {Avci}}, \bibinfo {author} {\bibfnamefont {O.}~\bibnamefont {Chmaissem}},
  \bibinfo {author} {\bibfnamefont {D.~Y.}\ \bibnamefont {Chung}}, \bibinfo
  {author} {\bibfnamefont {S.}~\bibnamefont {Rosenkranz}}, \bibinfo {author}
  {\bibfnamefont {E.~A.}\ \bibnamefont {Goremychkin}}, \bibinfo {author}
  {\bibfnamefont {J.~P.}\ \bibnamefont {Castellan}}, \bibinfo {author}
  {\bibfnamefont {I.~S.}\ \bibnamefont {Todorov}}, \bibinfo {author}
  {\bibfnamefont {J.~A.}\ \bibnamefont {Schlueter}}, \bibinfo {author}
  {\bibfnamefont {H.}~\bibnamefont {Claus}}, \bibinfo {author} {\bibfnamefont
  {A.}~\bibnamefont {Daoud-Aladine}}, \bibinfo {author} {\bibfnamefont {D.~D.}\
  \bibnamefont {Khalyavin}}, \bibinfo {author} {\bibfnamefont {M.~G.}\
  \bibnamefont {Kanatzidis}}, \ and\ \bibinfo {author} {\bibfnamefont
  {R.}~\bibnamefont {Osborn}},\ }\href {\doibase 10.1103/PhysRevB.85.184507}
  {\bibfield  {journal} {\bibinfo  {journal} {Phys. Rev. B}\ }\textbf {\bibinfo
  {volume} {85}},\ \bibinfo {pages} {184507} (\bibinfo {year}
  {2012})}\BibitemShut {NoStop}%
\bibitem [{\citenamefont {Avci}\ \emph {et~al.}(2013)\citenamefont {Avci},
  \citenamefont {Allred}, \citenamefont {Chmaissem}, \citenamefont {Chung},
  \citenamefont {Rosenkranz}, \citenamefont {Schlueter}, \citenamefont {Claus},
  \citenamefont {Daoud-Aladine}, \citenamefont {Khalyavin}, \citenamefont
  {Manuel}, \citenamefont {Llobet}, \citenamefont {Suchomel}, \citenamefont
  {Kanatzidis},\ and\ \citenamefont {Osborn}}]{Avci2013}%
  \BibitemOpen
  \bibfield  {author} {\bibinfo {author} {\bibfnamefont {S.}~\bibnamefont
  {Avci}}, \bibinfo {author} {\bibfnamefont {J.~M.}\ \bibnamefont {Allred}},
  \bibinfo {author} {\bibfnamefont {O.}~\bibnamefont {Chmaissem}}, \bibinfo
  {author} {\bibfnamefont {D.~Y.}\ \bibnamefont {Chung}}, \bibinfo {author}
  {\bibfnamefont {S.}~\bibnamefont {Rosenkranz}}, \bibinfo {author}
  {\bibfnamefont {J.~A.}\ \bibnamefont {Schlueter}}, \bibinfo {author}
  {\bibfnamefont {H.}~\bibnamefont {Claus}}, \bibinfo {author} {\bibfnamefont
  {A.}~\bibnamefont {Daoud-Aladine}}, \bibinfo {author} {\bibfnamefont {D.~D.}\
  \bibnamefont {Khalyavin}}, \bibinfo {author} {\bibfnamefont {P.}~\bibnamefont
  {Manuel}}, \bibinfo {author} {\bibfnamefont {A.}~\bibnamefont {Llobet}},
  \bibinfo {author} {\bibfnamefont {M.~R.}\ \bibnamefont {Suchomel}}, \bibinfo
  {author} {\bibfnamefont {M.~G.}\ \bibnamefont {Kanatzidis}}, \ and\ \bibinfo
  {author} {\bibfnamefont {R.}~\bibnamefont {Osborn}},\ }\href {\doibase
  10.1103/PhysRevB.88.094510} {\bibfield  {journal} {\bibinfo  {journal} {Phys.
  Rev. B}\ }\textbf {\bibinfo {volume} {88}},\ \bibinfo {pages} {094510}
  (\bibinfo {year} {2013})}\BibitemShut {NoStop}%
\bibitem [{\citenamefont {Barzykin}\ and\ \citenamefont
  {Gor'kov}(2009)}]{Barzykin2009}%
  \BibitemOpen
  \bibfield  {author} {\bibinfo {author} {\bibfnamefont {V.}~\bibnamefont
  {Barzykin}}\ and\ \bibinfo {author} {\bibfnamefont {L.~P.}\ \bibnamefont
  {Gor'kov}},\ }\href {\doibase 10.1103/PhysRevB.79.134510} {\bibfield
  {journal} {\bibinfo  {journal} {Phys. Rev. B}\ }\textbf {\bibinfo {volume}
  {79}},\ \bibinfo {pages} {134510} (\bibinfo {year} {2009})}\BibitemShut
  {NoStop}%
\bibitem [{\citenamefont {Qi}\ and\ \citenamefont {Xu}(2009)}]{Qi2009}%
  \BibitemOpen
  \bibfield  {author} {\bibinfo {author} {\bibfnamefont {Y.}~\bibnamefont
  {Qi}}\ and\ \bibinfo {author} {\bibfnamefont {C.}~\bibnamefont {Xu}},\ }\href
  {\doibase 10.1103/PhysRevB.80.094402} {\bibfield  {journal} {\bibinfo
  {journal} {Phys. Rev. B}\ }\textbf {\bibinfo {volume} {80}},\ \bibinfo
  {pages} {094402} (\bibinfo {year} {2009})}\BibitemShut {NoStop}%
\bibitem [{\citenamefont {Chubukov}\ \emph {et~al.}(2016)\citenamefont
  {Chubukov}, \citenamefont {Khodas},\ and\ \citenamefont
  {Fernandes}}]{Chubukov2016}%
  \BibitemOpen
  \bibfield  {author} {\bibinfo {author} {\bibfnamefont {A.~V.}\ \bibnamefont
  {Chubukov}}, \bibinfo {author} {\bibfnamefont {M.}~\bibnamefont {Khodas}}, \
  and\ \bibinfo {author} {\bibfnamefont {R.~M.}\ \bibnamefont {Fernandes}},\
  }\href {\doibase 10.1103/PhysRevX.6.041045} {\bibfield  {journal} {\bibinfo
  {journal} {Phys. Rev. X}\ }\textbf {\bibinfo {volume} {6}},\ \bibinfo {pages}
  {041045} (\bibinfo {year} {2016})}\BibitemShut {NoStop}%
\bibitem [{\citenamefont {Yu}\ and\ \citenamefont {Si}(2015)}]{Yu2015}%
  \BibitemOpen
  \bibfield  {author} {\bibinfo {author} {\bibfnamefont {R.}~\bibnamefont
  {Yu}}\ and\ \bibinfo {author} {\bibfnamefont {Q.}~\bibnamefont {Si}},\ }\href
  {\doibase 10.1103/PhysRevLett.115.116401} {\bibfield  {journal} {\bibinfo
  {journal} {Phys. Rev. Lett.}\ }\textbf {\bibinfo {volume} {115}},\ \bibinfo
  {pages} {116401} (\bibinfo {year} {2015})}\BibitemShut {NoStop}%
\bibitem [{\citenamefont {{Glasbrenner}}\ \emph {et~al.}(2015)\citenamefont
  {{Glasbrenner}}, \citenamefont {{Mazin}}, \citenamefont {{Jeschke}},
  \citenamefont {{Hirschfeld}}, \citenamefont {Fernandes},\ and\ \citenamefont
  {{Valent{\'{\i}}}}}]{Glasbrenner2015}%
  \BibitemOpen
  \bibfield  {author} {\bibinfo {author} {\bibfnamefont {J.~K.}\ \bibnamefont
  {{Glasbrenner}}}, \bibinfo {author} {\bibfnamefont {I.~I.}\ \bibnamefont
  {{Mazin}}}, \bibinfo {author} {\bibfnamefont {H.~O.}\ \bibnamefont
  {{Jeschke}}}, \bibinfo {author} {\bibfnamefont {P.~J.}\ \bibnamefont
  {{Hirschfeld}}}, \bibinfo {author} {\bibfnamefont {R.~M.}\ \bibnamefont
  {Fernandes}}, \ and\ \bibinfo {author} {\bibfnamefont {R.}~\bibnamefont
  {{Valent{\'{\i}}}}},\ }\href {\doibase 10.1038/nphys3434} {\bibfield
  {journal} {\bibinfo  {journal} {Nature Physics}\ }\textbf {\bibinfo {volume}
  {11}},\ \bibinfo {pages} {953} (\bibinfo {year} {2015})}\BibitemShut
  {NoStop}%
\bibitem [{\citenamefont {Scherer}\ \emph {et~al.}(2017)\citenamefont
  {Scherer}, \citenamefont {Jacko}, \citenamefont {Friedrich}, \citenamefont
  {\ifmmode \mbox{\c{S}}\else \c{S}\fi{}a\ifmmode \mbox{\c{s}}\else
  \c{s}\fi{}\ifmmode \imath \else \i \fi{}o\ifmmode~\breve{g}\else
  \u{g}\fi{}lu}, \citenamefont {Bl\"ugel}, \citenamefont {Valent\'{\i}},\ and\
  \citenamefont {Andersen}}]{Scherer_FeSe}%
  \BibitemOpen
  \bibfield  {author} {\bibinfo {author} {\bibfnamefont {D.~D.}\ \bibnamefont
  {Scherer}}, \bibinfo {author} {\bibfnamefont {A.~C.}\ \bibnamefont {Jacko}},
  \bibinfo {author} {\bibfnamefont {C.}~\bibnamefont {Friedrich}}, \bibinfo
  {author} {\bibfnamefont {E.}~\bibnamefont {\ifmmode \mbox{\c{S}}\else
  \c{S}\fi{}a\ifmmode \mbox{\c{s}}\else \c{s}\fi{}\ifmmode \imath \else \i
  \fi{}o\ifmmode~\breve{g}\else \u{g}\fi{}lu}}, \bibinfo {author}
  {\bibfnamefont {S.}~\bibnamefont {Bl\"ugel}}, \bibinfo {author}
  {\bibfnamefont {R.}~\bibnamefont {Valent\'{\i}}}, \ and\ \bibinfo {author}
  {\bibfnamefont {B.~M.}\ \bibnamefont {Andersen}},\ }\href {\doibase
  10.1103/PhysRevB.95.094504} {\bibfield  {journal} {\bibinfo  {journal} {Phys.
  Rev. B}\ }\textbf {\bibinfo {volume} {95}},\ \bibinfo {pages} {094504}
  (\bibinfo {year} {2017})}\BibitemShut {NoStop}%
\bibitem [{\citenamefont {Ishizuka}\ \emph {et~al.}(2018)\citenamefont
  {Ishizuka}, \citenamefont {Yamada}, \citenamefont {Yanagi},\ and\
  \citenamefont {{\=O}no}}]{Ishizuka2017}%
  \BibitemOpen
  \bibfield  {author} {\bibinfo {author} {\bibfnamefont {J.}~\bibnamefont
  {Ishizuka}}, \bibinfo {author} {\bibfnamefont {T.}~\bibnamefont {Yamada}},
  \bibinfo {author} {\bibfnamefont {Y.}~\bibnamefont {Yanagi}}, \ and\ \bibinfo
  {author} {\bibfnamefont {Y.}~\bibnamefont {{\=O}no}},\ }\href {\doibase
  10.7566/JPSJ.87.014705} {\bibfield  {journal} {\bibinfo  {journal} {Journal
  of the Physical Society of Japan}\ }\textbf {\bibinfo {volume} {87}},\
  \bibinfo {pages} {014705} (\bibinfo {year} {2018})}\BibitemShut {NoStop}%
\bibitem [{\citenamefont {Yip}\ \emph {et~al.}(2017)\citenamefont {Yip},
  \citenamefont {Chan}, \citenamefont {Niu}, \citenamefont {Matsuura},
  \citenamefont {Mizukami}, \citenamefont {Kasahara}, \citenamefont {Matsuda},
  \citenamefont {Shibauchi},\ and\ \citenamefont {Goh}}]{Yip2017}%
  \BibitemOpen
  \bibfield  {author} {\bibinfo {author} {\bibfnamefont {K.~Y.}\ \bibnamefont
  {Yip}}, \bibinfo {author} {\bibfnamefont {Y.~C.}\ \bibnamefont {Chan}},
  \bibinfo {author} {\bibfnamefont {Q.}~\bibnamefont {Niu}}, \bibinfo {author}
  {\bibfnamefont {K.}~\bibnamefont {Matsuura}}, \bibinfo {author}
  {\bibfnamefont {Y.}~\bibnamefont {Mizukami}}, \bibinfo {author}
  {\bibfnamefont {S.}~\bibnamefont {Kasahara}}, \bibinfo {author}
  {\bibfnamefont {Y.}~\bibnamefont {Matsuda}}, \bibinfo {author} {\bibfnamefont
  {T.}~\bibnamefont {Shibauchi}}, \ and\ \bibinfo {author} {\bibfnamefont
  {S.~K.}\ \bibnamefont {Goh}},\ }\href {\doibase 10.1103/PhysRevB.96.020502}
  {\bibfield  {journal} {\bibinfo  {journal} {Phys. Rev. B}\ }\textbf {\bibinfo
  {volume} {96}},\ \bibinfo {pages} {020502} (\bibinfo {year}
  {2017})}\BibitemShut {NoStop}%
\bibitem [{\citenamefont {{Lebert}}\ \emph {et~al.}(2017)\citenamefont
  {{Lebert}}, \citenamefont {{Bal{\'e}dent}}, \citenamefont {{Toulemonde}},
  \citenamefont {{Ablett}}, \citenamefont {{Klotz}}, \citenamefont {{Hansen}},
  \citenamefont {{Rodi{\`e}re}}, \citenamefont {{Raba}},\ and\ \citenamefont
  {{Rueff}}}]{Lebert2017}%
  \BibitemOpen
  \bibfield  {author} {\bibinfo {author} {\bibfnamefont {B.~W.}\ \bibnamefont
  {{Lebert}}}, \bibinfo {author} {\bibfnamefont {V.}~\bibnamefont
  {{Bal{\'e}dent}}}, \bibinfo {author} {\bibfnamefont {P.}~\bibnamefont
  {{Toulemonde}}}, \bibinfo {author} {\bibfnamefont {J.~M.}\ \bibnamefont
  {{Ablett}}}, \bibinfo {author} {\bibfnamefont {S.}~\bibnamefont {{Klotz}}},
  \bibinfo {author} {\bibfnamefont {T.}~\bibnamefont {{Hansen}}}, \bibinfo
  {author} {\bibfnamefont {P.}~\bibnamefont {{Rodi{\`e}re}}}, \bibinfo {author}
  {\bibfnamefont {M.}~\bibnamefont {{Raba}}}, \ and\ \bibinfo {author}
  {\bibfnamefont {J.-P.}\ \bibnamefont {{Rueff}}},\ }\href
  {http://adsabs.harvard.edu/abs/2017arXiv170804805L} {\bibfield  {journal}
  {\bibinfo  {journal} {ArXiv e-prints}\ ,\ \bibinfo {pages} {1708.04805}}
  (\bibinfo {year} {2017})}\BibitemShut {NoStop}%
\end{thebibliography}%

\end{document}